\documentclass[%
 reprint,
 amsmath,amssymb,
 aps,
]{revtex4-2}

\usepackage{graphicx}
\usepackage{dcolumn}
\usepackage{bm}
\usepackage{hyperref}

\usepackage{tikz}
\usetikzlibrary{arrows,backgrounds,snakes,patterns}
\usetikzlibrary{shapes,arrows,chains}
\usepackage{verbatim}
\usepackage{booktabs}
\usepackage{pgfplots}
\pgfplotsset{compat=1.10}
\usepgfplotslibrary{fillbetween}
\usepackage[flushleft]{threeparttable}
\graphicspath{ {./figures/} }
\usepackage{array}
\usepackage{placeins}

\usepackage{subcaption}
\usepackage{datetime2}
\usepackage{physics}
\usepackage{xspace}
\usepackage{amsmath}
\usepackage{amssymb}
\usepackage{amsfonts}

\newcommand{\glass}{\texttt{GLASS}\xspace}
\newcommand{\glassvi}{\texttt{GLASSv1}\xspace}

\newcommand{\glassviii}{\texttt{GLASSv3}\xspace}
\newcommand{\sangria}{\textit{Sangria}\xspace}
\newcommand{\radler}{\textit{Radler}\xspace}
\newcommand{\erebor}{\texttt{Erebor}\xspace}

\begin{document}


\title{Prototype Stochastic Gravitational Wave Background Recovery\\ in the LISA Global Fit Residual}

\author{Robert Rosati}
\email{robert.j.rosati@nasa.gov}
\altaffiliation[NASA Postdoctoral Program Fellow]{}
\author{Tyson B. Littenberg}%
 \email{tyson.b.littenberg@nasa.gov}
\affiliation{%
 	NASA Marshall Space Flight Center, Huntsville, AL 35812, USA
}%

\date{\today}

\begin{abstract}
The Laser Interferometer Space Antenna (LISA) mission poses a difficult parameter estimation challenge: the sources will be so dense in both time and frequency that they all must be fit simultaneously in a `global fit'.
Successful tests of global fit efforts on synthetic datasets have been recently reported, recovering extra-galactic black hole mergers and galactic binaries, including the \glass pipeline in \cite{Littenberg:2023xpl}.
Injected stochastic sources, however, have so far been absent in these datasets.
In this work we report our development of a stochastic search pipeline ready for inclusion in future tests of the global fit, capable of detecting or placing limits on a wide variety of possible cosmologically- and astrophysically-inspired SGWBs.
The code uses short-time Fourier transforms (STFTs) to allow for inference despite the non-stationarity of the noise.
We quote results using both purely synthetic confusion noise and two \glass residuals, and quantify the impact of the residuals' non-gaussianity on injected signal recovery and on setting upper limits. We find that, if not properly mitigated, non-gaussianities can preclude setting accurate SGWB upper limits and lead to false detections.
We also stress that the narrow-band non-gaussianities we find do not affect all sources equally, and many narrower-band, cosmologically-inspired SGWBs are more sensitive to non-gaussianity than others.
\end{abstract}

\maketitle


\section{\label{sec:intro}Introduction}

The Laser Interferometer Space Antenna (LISA) will open a new window of the gravitational wave spectrum, the milli-Hertz band. This region is expected to be extremely signal dense, with LISA capable of resolving thousands of binary systems in our own galaxy and tens to hundreds of extra-galactic black hole mergers.
Also present may be stochastic gravitational wave backgrounds, from sufficiently dense early- or late-universe sources.
Nearly all LISA sources will be so dense in time and frequency that parameter estimation, to remain unbiased, must be done for all sources simultaneously.
This procedure, called the global fit, is the fundamental challenge of LISA data analysis. It is a tour-de-force, requiring parameter estimation in a dauntingly large space, with an expected dimensionality of $\mathcal{O}(10^5)$ \cite{SciRD,redbook}.

Although this problem is difficult, it is not insurmountable, and global fit analyses have recently become available in the literature \cite{Littenberg:2023xpl,Strub:2024kbe,Katz:2024oqg} analyzing the \sangria LISA Data Challenge (LDC) \cite{le_jeune_2022_7132178}.
Each of them is based on a blocked Gibbs sampling scheme, where residuals are passed between a sampler for each source type.
This procedure has the advantage of allowing samplers to operate largely independently, and allows for the most highly correlated sampling (same source type) to occur within one block of the sampler.
So far, none of the prototype global analyses have included a dedicated stochastic background search (and there were no intrinsically stochastic sources injected into \sangria).
\begin{figure*}[th]
    \centering
    \includegraphics[width=0.95\textwidth]{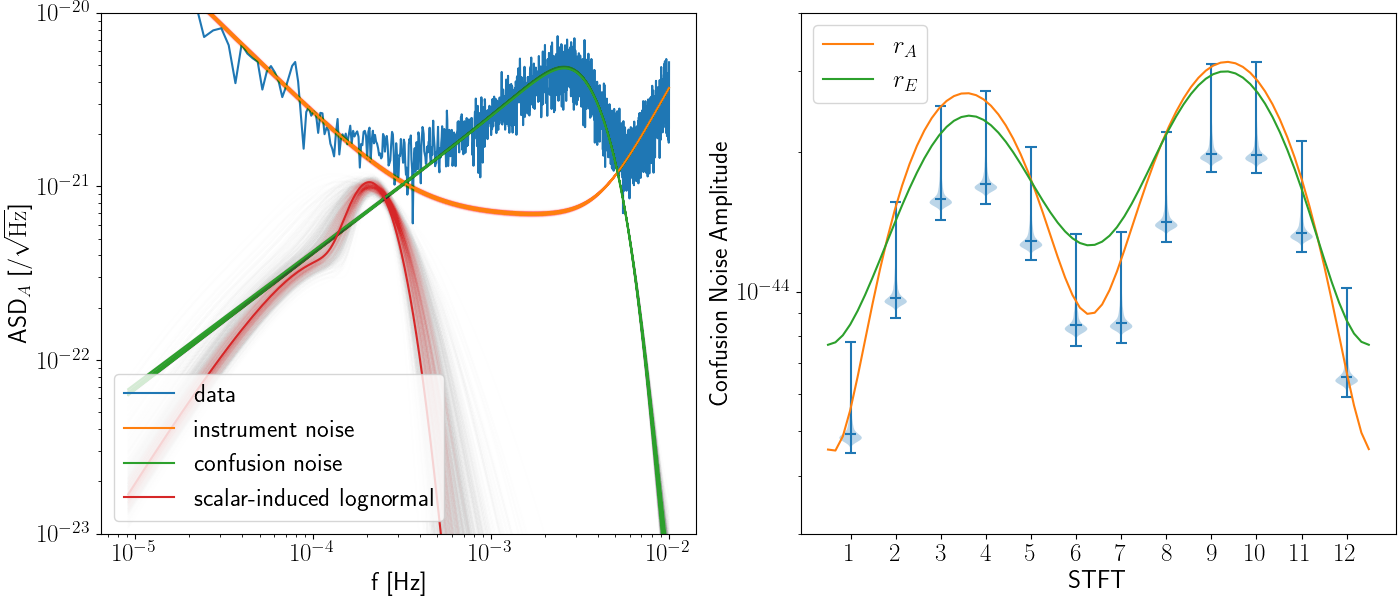}
    \caption{We demonstrate the ability of our pipeline to recover an injected cosmological background despite the presence of confusion noise. Left: we display the periodogram of the last month's data along with $10^3$ fair draws from the posterior, split into their corresponding stochastic components and plotted at low opacity. The injected values are plotted at full opacity.
    Right: we show the recovered amplitude of the confusion noise in each month-long short-time fourier transform. The injected amplitudes in the A and E channel are also plotted alongside the amplitude samples binned in violins.}
    \label{fig:lognormal_fit_violin}
\end{figure*}

At least one stochastic source is guaranteed to be in the LISA data, the unresolvable foreground of galactic binaries (GBs) left after the resolvable sources have been fit and subtracted from the data \cite{Cornish:2015pda}.
This so-called confusion noise will be anisotropic in the sky, and will form a time-varying source in the LISA data during the constellation's orbit (in other words, it is an intrinsically stationary source with extrinsic non-stationarity).
Several other possible stochastic sources include the stellar-origin black hole binary background (SOBHBs) \cite{Phinney:2001di,Buscicchio:2021dph,Babak:2023lro}, a background of extreme mass-ratio inspirals (EMRIs) \cite{Bonetti:2020jku}, or various possible early-universe dynamics including first-order phase transitions \cite{Caprini:2019egz,Caprini:2024hue}, the collapse of large primordial density perturbations from inflation \cite{Acquaviva:2002ud,Baumann:2007zm,Espinosa:2018eve,Kohri:2018awv,SIGWstub}, or cosmic strings \cite{Auclair:2019wcv,Blanco-Pillado:2024aca}.

Several stochastic searches for LISA exist in the literature, which differ in their treatment of the confusion noise, any other stochastic signals, and the instrument noise.
Perhaps the most agnostic option is to not distinguish between the instrumental noise and stochastic signals, assume the instruments' data streams are uncorrelated, and fit the residual in each with a non-parametric spline, allowed to vary arbitrarily in time and frequency.
This would encapsulate all stochastic sources and instrumental noise while allowing for non-stationarity, but would also minimally constrain what any stochastic sources are and use no assumptions about the instrument.
This might also be impossible in practice, as allowing arbitrary non-stationarity could also absorb even non-stochastic signals!
A stationary version of this procedure is however possible and is the approach taken by the Global LISA Analysis Software Suite (\glass) in ref. \cite{Littenberg:2023xpl}, the analysis we call \glassvi.

Another option is to fit either the instrument noise or the signals with parameterized models, strongly assuming their shape as a function of frequency.
This approach has the advantage of a lower dimensional parameter space, at the cost of more aggressive assumptions about the signal or instrument model, vulnerable to biases when these assumptions are inaccurate.

A parametric noise but non-parametric SGWB search has been implemented in \texttt{SGWBinner}, developed by members of the LISA Cosmology Working Group \cite{Caprini:2019pxz,Flauger:2020qyi}. \texttt{SGWBinner} assumes all sources are extrinsically stationary, and can be described by a piecewise-powerlaw fit.
An implementation of a fully parametrized instrument and signal model was in \cite{Adams:2013qma}, which used the galactic population and the LISA constellation's orbit to fix the confusion noise amplitude as a function of time. Similarly \cite{Digman:2022jmp,Hindmarsh:2024ttn} used a parametric signal and noise model in a mixed time-frequency basis and modeled the confusion noise as an isotropic, intrinsically cyclo-stationary signal.
The \texttt{BLIP} package in \cite{Banagiri:2021ovv, Mentasti:2023uyi} models stochastic signals as intrinsically stationary but anisotropic, and recovers their sky distributions from their extrinsic non-stationarity. Both works use a fully parametrized instrument and signal model, but consider a non-parametric sky localization with a spherical harmonic basis.

It is also possible to take into account likely uncertainties in the instrument's noise performance and attempt to recover stochastic signals anyway.
In Ref. \cite{Baghi:2023qnq}, the authors modeled the instrument at the laser link level and found that noise knowledge uncertainty did not adversely affect stochastic signal recovery; however Ref. \cite{Muratore:2023gxh}, under different assumptions, found that marginalizing over instrumental uncertainties caused an order of magnitude or worse degradation in SGWB upper limits.
Refs. \cite{Pozzoli:2023lgz,Alvey:2023npw} use machine-learned descriptions of the stochastic signals and noise that allow for additional correlations compared to an idealized instrument model.

In this work, we focus on the challenges of performing a stochastic search in a global fit pipeline and therefore model the instrument ideally, following the model used in \sangria \cite{le_jeune_2022_7132178} and more recent runs of \glass (though not in \cite{Littenberg:2023xpl}). We assume a parametrized, isotropic, and non-stationary signal model, and a simple two-parameter instrument noise model.
As a first in the literature, we work with true global fit residuals for the confusion noise.

An example successful recovery of an injected signal with synthetic confusion noise using our pipeline is visible in Figure \ref{fig:lognormal_fit_violin}.
On the left we show the periodogram of the last month of the data and several fair draws of the posterior, successfully reconstructing the injected SGWB, the confusion noise, and the instrumental noise.
On the right, we show how we account for the non-stationarity of the confusion noise. We split the year of data into several short time segments and fit its amplitude independently in each of the 12, which we show binned in violins compared to the injected confusion time dependence.

In this paper, we focus on the subtleties of applying these techniques to real global fit residuals.
They are non-gaussian at many frequencies, from either insufficient convergence of the MCMC or the `popcorn noise' regime where too few sources are contributing to the background for the central limit theorem to apply \cite{Romano:2016dpx}.
These non-gaussianities, if unmitigated, bias any stochastic signal inference and give false detections when attempting to set upper limits for different signal types.

In Section \ref{sec:residual}, we describe these data analysis challenges and in Section \ref{sec:implementation} our pipeline that allows for non-stationarity for the confusion noise and has mitigation measures for the non-gaussianities.
Later in Section \ref{sec:results}, we perform both detection and upper limit studies with our pipeline, comparing our performance on the \texttt{GLASS} residual to performance on a synthetic idealized confusion noise.
Finally in Section \ref{sec:conclusions} we summarize our results and discuss how we might improve future stochastic searches.

\section{The Global Fit Residual}
\label{sec:residual}

The principal challenge of LISA data analysis can be understood in terms of the Gaussian likelihood
\begin{equation}
\begin{aligned}
    \log \mathcal{L} = \Sigma_f\big[  &-\Sigma_{ij} \Delta d_i(t,f) (C^{-1}(t,f))_{ij} \Delta d_j(t,f)\\
    &- \log |C| -\log 2\pi  \big]
\end{aligned}
\label{eq:likelihood}
\end{equation}
where $\Delta d_i = d_{i,{\rm signal}} - d_{i,{\rm model}}$ and $i,j$ range over the available data channels. 
Our data analysis problem is essentially to determine the components of $d_{i,{\rm signal}} = d_{\rm GB} + d_{\rm MBHB} + \ldots$ so that the residual $\Delta d_i$ is approximately normally distributed, with a covariance given by $C_{ij}(t,f)$. 
$C_{ij}(t,f)$ includes all stochastic contributions to the data, including instrumental noise and any astrophysical or cosmological backgrounds.
We expect these stochastic contributions to be approximately gaussian, because many independent sources sum to either make the instrumental noise or stochastic signals. We revisit this assumption later in Section \ref{sec:pipeline-nongaussianity}.

Because the various terms in $d_{i,{\rm signal}}$ interfere, the necessity of a global solution to LISA parameter estimation has been long argued \cite{Cornish:2005qw}.
The LISA band will be too dense, in both time and frequency, to perform unbiased parameter estimation on individual sources.

It is ideal then, to perform any stochastic parameter inference during the global fit itself. This should not only improve the fit by separating out the instrumental noise and other, potentially non-stationary stochastic backgrounds, but also allow for improved parameter estimation for the stochastic sources by properly marginalizing over source uncertainties.

Although we do not achieve a global stochastic fit in this work, we develop a pipeline with the ability to do so in future global fit runs.
We use the residuals from the prototype global fit pipeline \glass applied to the \sangria LISA Data Challenge \cite{le_jeune_2022_7132178}. \glass and other groups developing prototype global fits have based them on a blocked Gibbs sampling scheme, where trans-dimensional Markov-chain Monte Carlo (MCMC) samplers for each source type trade turns sampling until a convergence criterion is reached \cite{Littenberg:2023xpl,Katz:2024oqg}.

At this point, \glass has produced posterior samples for all source types, including MBHBs, GBs, and a stochastic residual (assumed stationary).
We consider \glass runs with two possibilities for this residual, one analytic model described below (\glassviii), and one fit by an Akima spline (\glassvi).
These residuals either separate the confusion noise component (\glassviii) or fit a sum of all stochastic components (\glassvi).

The exact nature of the confusion noise in a global fit has not been previously explored in the literature.
In a trans-dimensional global fit pipeline, the confusion noise is set by the number of discrete templates accepted by the GB sampler, and how well these cancel the input waveforms.
One very naive and computationally impractical option would be to not include a ``confusion noise'' at all, and model the unresolvable GBs as discrete signals contributing to $d_{i,{\rm signal}}$.
Allowing for infinite computing resources, this sea of unresolved sources would accurately describe the confusion foreground.
But this scenario would introduce so many spurious parameters that it is better to marginalize over them, and treat the unresolved background as a contribution to the noise covariance $C$.
Ref. \cite{Cornish:2015pda} describes this derivation analytically and shows that the combination of some discrete sources and a parametrized confusion noise is the right model in a Bayesian sense -- it has a higher evidence for most populations compared to discrete-only source models or a non-gaussian confusion noise model.
As we describe below, the spectral shape of the confusion noise from a given population is difficult to predict when including realistic source-to-source covariances.
Similarly, the confusion noise has been assumed to be gaussian at all frequencies, and it is under the assumptions of \cite{Cornish:2015pda} with a dense enough population, but the galactic foreground may not meet these criteria at the high-frequency boundary of the confusion noise, depending on the realization of the galactic population.
It may also be the case that fully-converged trans-dimensional MCMC can only leave a gaussian confusion noise.

It also is difficult to know how well the current prototype global fits are fitting the unresolved source population.
So far in the literature, \cite{Littenberg:2023xpl,Katz:2024oqg} have focused their performance metrics (e.g. the match) and cataloging steps on the reported high-confidence sources, which allows them to directly compare their results to the injected populations.
Unfortunately these metrics have only a tangential relationship to the confusion noise, which is very sensitive to the low-amplitude and sub-threshold sources that do not make it into the catalog.

Let's consider two possible fits to the GB spectrum: one that includes some low-SNR sources with poorly constrained or unconstrained parameters, and one that does not. The cataloging process would find identical high-SNR sources and therefore make identical GB catalogs, but \textit{the case with sub-threshold sources would subtract more power out of the residual, making it more gaussian and avoiding false detections during the SGWB search}.
It is one of the principal results of this paper that global fits should not only be judged by how well they recover resolvable sources, but also by the gaussianity of and total power left in their residuals.
It it entirely possible to recover all high-SNR sources well but mis-estimate the optimal confusion noise.
Of course, more sub-threshold sources are not always better for the residual, and an overparametrized fit to the unresolved population may rob power from other signals.

For the analysis in this work, we use a single fair draw from each of the \glass residuals' posteriors.
Working with only one sample of the residual, the analysis described here is not a true global stochastic analysis, which would require the stochastic sampler to perform parameter estimation live, during the Gibbs scheme.
However by passing a fair draw of the stochastic parameters to the other samplers, getting a new residual from them in return, and repeating the process $\gtrsim \mathcal{O}(10^3)$ times, the pipeline we describe below could be used to perform true global stochastic parameter estimation.
We plan to do just this during future \glass runs.


\begin{figure*}[t!]
    \centering
\includegraphics[width=\textwidth]{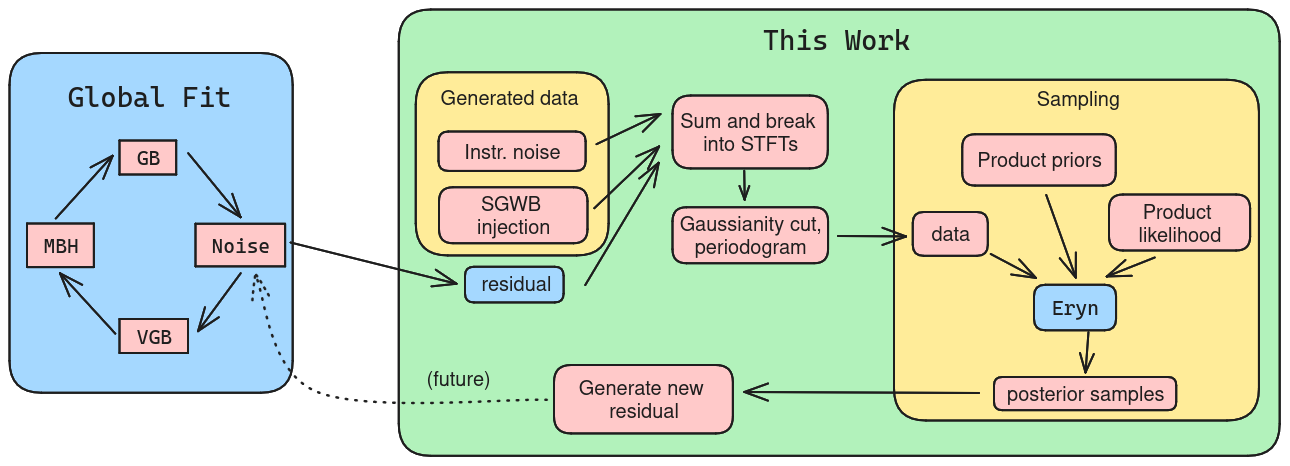}
    \caption{The architecture of the SGWB recovery pipeline presented in this work.}
\label{fig:architecture}
\end{figure*}

Our pipeline's architecture diagram is available in Figure \ref{fig:architecture}.
In order to recover any stochastic signals in the LISA data, we need to estimate the covariance matrix $C(t,f)_{ij}$ in the likelihood and decompose it into each stochastic contribution: the instrumental noise, the galactic foreground, and any extragalactic or cosmological backgrounds.
To prepare the data, we first take a global fit residual, subtract the \sangria noise because it does not match its published model, and add our own generated noise and any stochastic signal injections using the procedure in Sec \ref{sec:data-generation}.

To handle the non-stationarity of the galactic foreground, we divide the data into short time segments and periodogram each one. We then take a product likelihood between the segments, re-using most parameters in each segment but allowing the confusion noise amplitude to be independent in each. We comment more on non-stationary inference in general and our choices in Section \ref{sec:pipeline-nonstationarity}.
The residual is also not gaussian at all frequencies -- we discuss our mitigation measures in Section \ref{sec:pipeline-nongaussianity}.
Finally we sample the noise and signal parameters using \texttt{Eryn} \cite{Karnesis:2023ras}, and discuss our sampler configuration in Section \ref{sec:pipeline-sampling}.

We have implemented these processes in a proof-of-concept Python-based pipeline, with the core likelihood and SGWB template library written in C and called from Python, with the goal of them being accessible to the C-language \glass architecture in the future.

\section{Implementation}
\label{sec:implementation}

As we mentioned in Section \ref{sec:residual}, the challenge of stochastic signal searches in the LISA data is equivalent to estimating the residual's covariance matrix in \eqref{eq:likelihood}, and properly decomposing it into its constituent contributions.
To accomplish this, we match \sangria and assume that LISA's armlengths and the link noise levels are all equal, and write the covariance matrix for the three TDI channels $A,E,T$ \cite{Vallisneri:2004bn} as
\begin{equation}
\begin{aligned}
    C(t,f) = \mathrm{diag}( &N_A(f) + S_h(t,f) R_A(t,f),\\
    &N_E(f) + S_h(t,f) R_E(t,f),\\
    &N_T(f) + S_h(t,f) R_T(t,f)),
\end{aligned}
\end{equation}
where $N_i(f)$ are the noise model, $S_h(t,f)$ is a sum of all (potentially non-stationary) stochastic signal components, and $R_i(t,f)$ are the appropriate response functions. In principle these depend on the constellation's and sources' relative orientations and sky distributions.

In this work we make the simplifying assumption that all stochastic sources are intrinsically stationary and only vary in amplitude due to the constellation's motion (i.e. they are extrinsically non-stationary).
It is possible to use this amplitude variation and the knowledge of the $R_i(t,f)$ during the constellation's orbit to recover the stochastic source's anisotropy, as this is the approach taken by the \texttt{BLIP} package \cite{Banagiri:2021ovv,Rieck:2023pej}.
The constellation's motion also induces a small kinetic dipole and quadrupole anisotropy to SGWBs, which we will ignore \cite{Heisenberg:2024var}.

We instead model the extrinsic non-stationarity due to anisotropy as intrinsic non-stationarity in some stochastic parameters, freezing each channel's response in time, so that $S_h(t,f) R_i(t,f) \rightarrow S_h(f,\theta(t)) R_i(f)$.
We assume the response functions are equal to their sky- and polarization-averages at all times, which we take from \cite{Hartwig:2023pft}.

\subsection{Data}
\label{sec:data-generation}


To gradually approach the challenge of working with global fit pipeline residuals, we take several different options for the injected confusion noise, attempting to isolate the effects of non-stationarity and non-gaussianity.
We consider no confusion noise at all, synthetic non-stationary confusion noise, and finally residuals from two \texttt{GLASS} runs.
\glassvi, the residual from the results quoted in \cite{Littenberg:2023xpl}, used a flexible noise model based on an Akima spline, independent in the A and E channels and ignored the T channel.
\glassviii, a more recent re-run of \glass with several architectural improvements, used a parameterized noise model similar to \eqref{eq:noise} but with independent link-level amplitudes, as well as a confusion noise parametrization similar to \eqref{eq:confusion}.
\glassviii also used all three data channels, internally sampling in the $X, Y, Z$ TDI basis, which we have converted to $A, E, T$.

For now, we both inject and reconstruct any stochastic signal in the data with parametrized models.
These have the advantage of giving an increase in sensitivity when the signal is of a known shape, but of course require a strong expectation of the possible signal types.
We have compiled a small library of $\sim 20$ possible SGWB spectral shapes, both from early-universe cosmology and astrophysics. In this work, we present results using three representatives: a powerlaw (common in many astrophysically sourced backgrounds), a templated early-universe first-order phase transition (from beyond standard model physics in the electroweak sector or many other speculative early-universe scenarios), and the background that would be sourced by a primordial overdensity log-normal in wavenumber (common in many primordial black hole and inflationary models).

From a data analysis perspective, these have the nice property of encapsulating three stochastic signal bandwidths, with the powerlaw the broadest and the log-normal overdensity the narrowest.
As we discuss in Section \ref{sec:results}, these test the local gaussianity of the galactic foreground at different levels: the broadband backgrounds are the least sensitive to confusion noise mismodeling, while the narrow-band backgrounds are the most demanding of the residual.

We take an identical powerlaw parametrization as in the \radler LDC1-6\footnote{\url{https://lisa-ldc.lal.in2p3.fr/challenge1}}:
\begin{align}
    \Omega_{\mathrm{GW}}(f) = A_p \left( \frac{f}{25 \mathrm{ Hz}}\right)^{\alpha_p}
    \label{eq:powerlaw}
\end{align}
where we quote the background in terms of cosmological energy density, and the expected strain PSD is $S_h(f) = 3H_0^2/f^3 \times \Omega_{\mathrm{GW}}(f)$, and we take $H_0 = 70\, \mathrm{km/s/Mpc}$.
For the cases when we attempt a recovery of this signal, we take as our fiducial parameters the same amplitude taken in \radler LDC1-6: $\log_{10} A_p = -8.45, \alpha_p = 2/3$.

The collisions of domain walls, turbulence, and sound waves during first order phase transitions in the early universe can generate a complex stochastic gravitational wave background from the large variations in energy density during the phase transition.
In high fidelity simulations over a wide range of parameter space (at least for the so-called weak transitions), the sound waves have been found to be the dominant effect. These have been shown to combine into an approximately double-broken powerlaw SGWB, which can be parameterized by \cite{Boileau:2022ter}
\begin{equation}
\begin{aligned}
    \Omega_{\rm GW}(f) &= A_{\rm ph} s^9 \left(\frac{1+r_b^4}{r_b^4 + s^4}\right)^{(9-b)/4} \\
    &\times \left(\frac{b+4}{b+4 - m + m s^2} \right)^{(b+4)/2} \\
    m &= \frac{9r_b^4 + b}{r_b^4+1} \\
    s &= f/f_{\rm ph}
\end{aligned}
\label{eq:pt}
\end{equation}
where $f_{\rm ph}$ is the location of the peak, $r_b$ is the ratio between the two breaks in the spectrum, $b$ models the slope between the two breaks.
Note that other templates more accurately modeling strong transitions by considering turbulence and bubble collisions are now available \cite{Caprini:2024hue}, but we consider them beyond the scope of the present work.

The final shape of background we've chosen to study in this work are the gravitational waves formed when a log-normal scalar overdensity collapses during the radiation era of the early universe. This is a generic and model-independent signal, and could be generated from any early-universe dynamics that source large scalar perturbations to generate these scalar-induced gravitational waves \cite{Baumann:2007zm,Kohri:2018awv}.
The template is formed from a log-normal in the primordial density perturbation $P_\mathcal{R}(k)$, which can be written:
\begin{align*}
P_\mathcal{R}(k) &= A_\mathcal{R} \exp \left[-\frac{1}{2} \left( \frac{\log{k/k_\star}}{\Delta}\right)^2 \right] \\
\Omega_\mathrm{GW}\left(k = \frac{2 \pi f}{c} \right) &= \int T_\mathrm{RD}(x,y) P_\mathcal{R}(xk) P_\mathcal{R}(yk) \dd{x} \dd{y}
\end{align*}
where the width of the log-normal is $\Delta$, the amplitude of the overdensity is $A_\mathcal{R}$, and the peak frequency is given by $f_{\star} = \frac{c}{2\pi} k_\star$.
The gravitational wave spectrum from the collapse of this overdensity is in general calculable by an integration against a transfer function $T_\mathrm{RD}$ taking into account the equation of state of the universe at the time and all subsequent propagation effects \cite{Baumann:2007zm}, and in this particular example the integral happens to be solvable analytically. For brevity we do not reproduce the lengthy expression here, but refer the reader to \cite{Pi:2020otn}.

We also parametrize the confusion noise with a template, originally from \cite{Cornish:2017vip}
\begin{align}
    S_h(t,f) = A f^{-7/3} e^{-f^\alpha + \beta f \sin{(\kappa f)}}(1+\tanh(\gamma (f_k - f))).
    \label{eq:confusion}
\end{align}
When we treat the confusion noise as a non-stationary source, we take only its amplitude to be a function of time $A=A(t)$, and we assume the shape parameters and knee frequency ($\alpha,\beta,\gamma,\kappa,f_k$) are stationary.
We verified this approximation in Appendix \ref{sec:confvartest}, and as can be seen in Figure \ref{fig:confvartest} these do not change significantly if given freedom to vary in each STFT, at least not using the \glass residuals.
These shape parameters will vary during an online analysis, as more GBs become resolvable with longer observation time. However, for a fixed time period of data, which sources are resolvable and unresolvable will not vary time segment to time segment.

We use the simple two-parameter instrument noise model given for \sangria \cite{le_jeune_2022_7132178}:
\begin{equation}
\begin{aligned}
S_{A,E}(f) = 8 \sin^2{(f/f_\star)} \big[&2 S_{pm} (3 + 2 \cos{(f/f_\star)} + \cos{(2f/f_\star)})\\
&+ S_{op} (2+\cos{(f/f_\star)}) \big] \\
S_T(f) = 16 \sin^2{(f/f_\star)} \big[ &8 S_{pm}\sin^4{(f/(2f_\star))}\\ &+ S_{op} (1-\cos{(f/f_\star)}) \big],
\end{aligned}
\label{eq:noise}
\end{equation}
where $f_\star=2\pi c / L$, $L = 2.5 \times 10^9 \textrm{ m}$ is the constellation armlength, and the acceleration and optical metrology noises are given by
\begin{equation}
    \begin{aligned}
        S_{pm} = S_a &\left(1 + \left( \frac{4\times 10^{-4} \textrm{ Hz}}{f} \right)^2 \right) \left( 1 + \left(\frac{f}{8\times 10^{-3} \textrm{ Hz}}\right)^4 \right)\\
        &\times \left(\frac{1}{2\pi f}\right)^4 \left(\frac{2\pi f}{c}\right)^2
        \\
        S_{op} = S_i &\left( 1+\left(\frac{2\times 10^{-3} \textrm{ Hz}}{f} \right)^2 \right)\left(\frac{2\pi f}{c}\right)^2
    \end{aligned}
\end{equation}
Because the \sangria noise does not fit its published model \eqref{eq:noise} at low frequencies in the $T$ channel, we generate our own noise.

Referencing our architecture diagram (Figure \ref{fig:architecture}), to generate the data for a particular injection we create a timeseries from each desired contribution to the data and sum them. For example, for a powerlaw injection with \glassvi confusion noise, we would generate independent 3-channel timeseries for the instrumental noise and the powerlaw SGWB, and then add these to the \sangria-noise subtracted \glassvi residual.
How we generate a timeseries from a given desired powerspectral density (PSD) we leave to Appendix \ref{sec:noise_generation}.

\subsection{Non-stationarity}
\label{sec:pipeline-nonstationarity}

\begin{figure}[t]
    \centering
    \includegraphics[width=0.9\columnwidth]{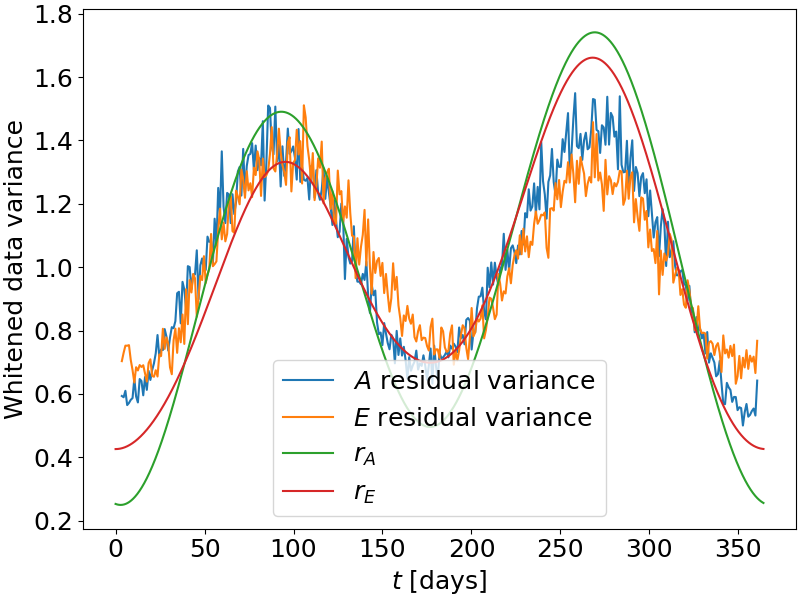}
    \caption{The variance of 1-day segments of the noise-free \glassvi residual (normalized by the average variance) compared to $r_A(t), r_E(t)$, an estimate of the confusion noise amplitude in the same dataset by \cite{Digman:2022jmp}.}
    \label{fig:rae_comparison}
\end{figure}

Stochastic sources in the LISA data have the possibility of being non-stationary (time-varying PSD), either extrinsically from source anisotropy on the sky, induced by instrumental sensitivity changes, or intrinsically non-stationary signals.

The loudest stochastic source in the LISA band is expected to be the galactic binary foreground, which will be extrinsically non-stationary. We plot the measured non-stationarity in the \glass residual as compared to an estimate from the literature in Figure \ref{fig:rae_comparison}. We show the normalized variance of day-long time segments of the residual in the A and E channels, which qualitatively matches a fit to the confusion noise time dependence in \cite{Digman:2022jmp}. The \glassvi residual does not perfectly match the parametric fit -- this could be because the fit determines which sources belong to the confusion noise based on an iterative SNR cut scheme rather than a method that takes into account source-to-source correlations like RJMCMC.

Other SGWB sources may also have some non-stationarity.
Extra-galactic astrophysical backgrounds are expected to have small anisotropies due to source concentrations in galaxies rather than voids, and therefore have some extrinsic non-stationarity. Primordial SGWBs are typically expected to have very small intrinsic anisotropies \cite{LISACosmologyWorkingGroup:2022kbp} and be mostly indistinguishable from isotropic sources in LISA, with up to percent-level kinematic anisotropies for the loudest possible sources \cite{Heisenberg:2024var}.

Some LISA stochastic sources may also be intrinsically non-stationary, with the main example being the `popcorn noise' effect of some transient sources dominating over the rest of a background formed by source confusion. This popcorn noise can also lead to non-gaussianity, as recently explored in \cite{Buscicchio:2024wwm,Piarulli:2024yhj,KarnesisStub}.

We have chosen to model non-stationarity by assuming an isotropic instrument response and that the signal is piecewise-stationary,  dividing the data into several short-time Fourier transforms. 
For this proof-of-concept work, we take an agnostic approach and do not enforce any relation of the non-stationary parameters to themselves in the different time segments (the remaining stationary parameters are of course assumed to be identical in each time segment). In practice our implemented product likelihood takes the form

\begin{equation}
\begin{aligned}
\log{\mathcal{L}} &= \sum_{i\in \mathrm{STFTs}} \log{\mathcal{L}_i}(\theta_{\mathrm{s}}, \theta_{\mathrm{ns},i}), \\
\log{\mathcal{L}_i} &= -\frac{N_{\mathrm{segs}}}{2} \left( \log{2\pi} + \sum_{j \in \mathrm{A,E,T}}\frac{\mathrm{PSD}_j}{C_j}  + \log{C_j} \right),
\end{aligned}
\label{eq:likelihood_implementation}
\end{equation}
where $N_{\mathrm{segs}}$ is how many data segments are internally averaged over during periodogramming, $\theta_{\mathrm{s}}$ and $\theta_{\mathrm{ns},i}$ are the stationary and non-stationary parameters in time segment $i$, and there is also an implicit sum over the (suppressed) frequency bin index.
In our case $N_\mathrm{segs}$ is not extremely high, and this may create some bias.

\begin{figure*}[t]
    \centering
    \includegraphics[width=0.49\textwidth]{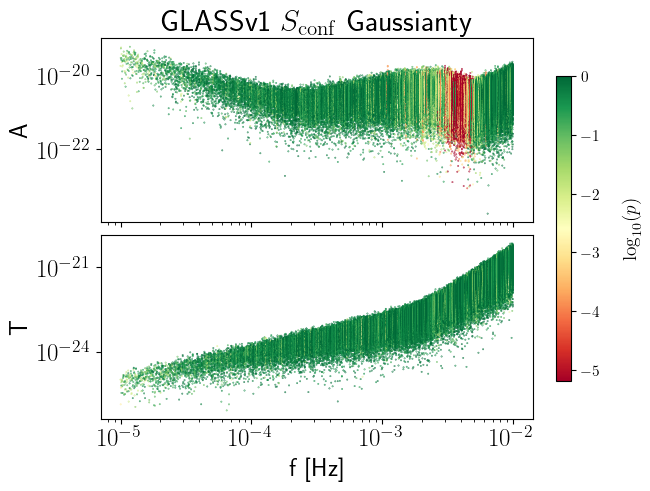}
    \includegraphics[width=0.47\textwidth]{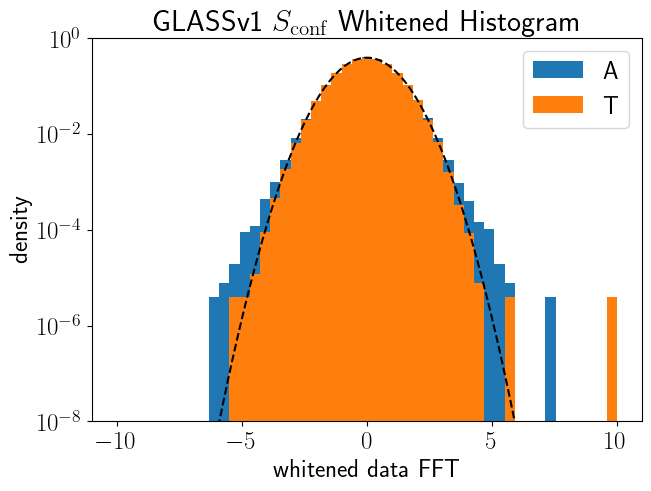}\\
    \includegraphics[width=0.49\textwidth]{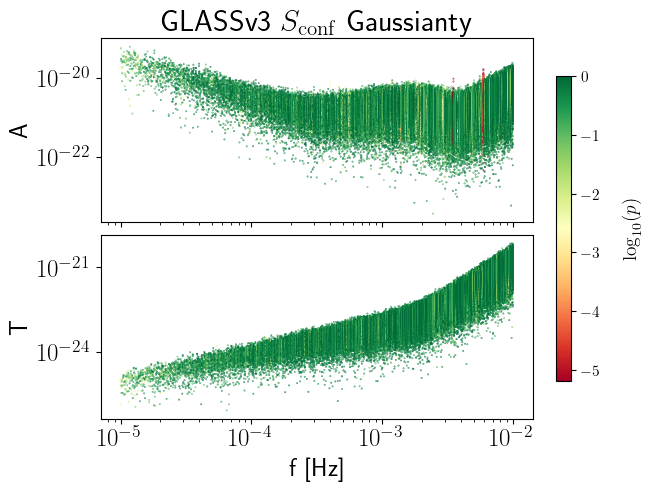}
    \includegraphics[width=0.47\textwidth]{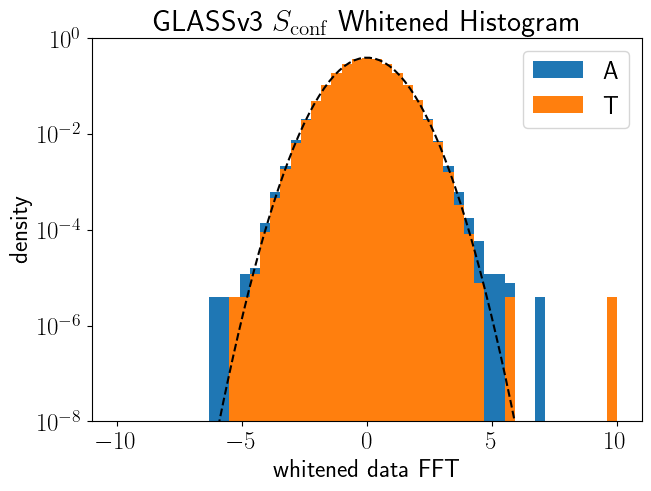}
    \caption{On the left, we show the amplitude spectral density of the A and T channels of datasets using only \glass confusion noise and instrumental noise, without an injected cosmological SGWB.
    The data are combined into 1548 bins, with boundaries equally log-spaced in frequency. In each bin, we whiten the data and perform an Anderson-Darling gaussianity test on the real and imaginary parts. The ASD plot is then colored by the log of the $p$-value of being drawn from a Gaussian distribution, which we cut off at $p=10^{-2}/N_{\rm bins}$. Significant non-gaussianity is present in the \glassvi confusion noise around 2-7 mHz, in the region with the highest density of resolvable sources. Note that the $p$-values have been cropped for readability, the lowest bin has a $p$-value of $\sim 10^{-24}$.
    The \glassviii confusion noise is markedly improved, with only a few non-gaussian bins. Its least gaussian bin has a $p$-value of $\sim 10^{-13}$.
    In the right column we study the form that the non-gaussianity takes, by histogramming the whitened real and imaginary parts of the entire frequency spectrum. In both \glass residuals, the gaussianity takes the form of heavy tails, with \glassvi showing departures from gaussianity (dashed black line) much earlier than \glassviii.}
    \label{fig:residual_gaussianity}
\end{figure*}

This is equivalent to complete agnosticism about the periodicity of the signals and allows for arbitrary non-stationarity of any source or instrumental noise parameter, but also minimally constrains the source through its time dependence. Assuming more information about the signal (e.g. that it is cyclo-stationary or that it comes from an intrinsically stationary source with a fixed anisotropic sky distribution) will decrease the posterior volume.
Because we do not make these assumptions, the results we report are conservative in this regard. In future work, we will use the predictable non-stationarity of the galactic foreground to better constrain it.

We also attempted to use a gaussian mixture model approach for this non-stationarity which we eventually abandoned. The idea was to model the posterior of one time segment as the prior for the subsequent time segment in all stationary parameters, but letting non-stationary parameters be drawn from their original priors. While a perfect reproduction of the posterior would have worked, we found too much error in the gaussian mixture model reconstruction of the posterior in all but the most trivial cases. We document this attempt in Appendix \ref{sec:GMMs}.

\subsection{Non-gaussianity}
\label{sec:pipeline-nongaussianity}

In gravitational wave data analysis, stochastic signals are often assumed to be a contribution of sufficiently many independent and identically distributed (i.i.d.) sources that the central limit theorem applies in every $(t,f)$ bin, so that the gaussian likelihood \eqref{eq:likelihood} remains an accurate description of the data \cite{Romano:2016dpx}.

Either or both of these assumptions may fail and generate non-gaussianity in the data: the sources which have been superimposed into one bin may not be i.i.d. and have correlations (e.g., due to non-stationarity or a common signal model), or there may be too few i.i.d. sources in a bin to ensure the central limit theorem is accurate, the so-called `popcorn noise' regime.

We find significant non-gaussianity in our global fit residuals, with some contributions likely from both of the above effects.
In Figure \ref{fig:residual_gaussianity}, in the left column, we group small windows in frequency and use a gaussianity measure based on the Anderson-Darling score \cite{10.1214/aoms/1177729437} to compute the $p$-value of each bin being samples drawn from a gaussian distribution (after whitening).
\glassvi shows several bins at $\sim 4$ mHz with significant non-gaussianity (the lowest $p$-value is $10^{-24}$), while the situation is improved in \glassviii, with few visible bins saturating the scale.
In the right column we show the histogram of the whitened \glassvi and \glassviii residuals. The non-gaussianity takes the form of significantly heavier tails than a gaussian distribution, with the departures from gaussianity significant even for \glassviii.

There are several possible causes for this non-gaussianity.
The non-stationarity of the confusion noise should in principle create correlations which induce some non-gaussianity\footnote{If we assume an otherwise stationary background $S_h(t)$ has some modulation $r(t)$, then its Fourier domain representation would be the convolution of the two functions:  $S_h(t)\times r(t) \Rightarrow \tilde{S}_h(f) \otimes \tilde{r}(f)$. A non-stationary background does not have independent frequency bins, and they are correlated at a bandwidth set by the modulation ($\sim 1/T$ for a sinusoidal modulation, where $T$ is the observation time).}, but if these were large they should be present in the entire confusion-noise-dominated portion of the data. Our observed non-gaussianities do not follow this pattern (and also \glassvi and \glassviii have starkly different gaussianities despite both using the \sangria data), so this effect cannot be the dominant one.

A popcorn noise effect could also be possible, as the region around $\sim 4$ mHz is when the GB population density has decreased drastically. At least in an old estimate (\cite{Timpano:2005gm}), the GB population decreases below $\sim 10$ sources per frequency bin at $>1$ mHz.
But it is not obvious to us that popcorn noise is a true effect in (perfect) trans-dimensional MCMC. A sum of low-amplitude discrete signals can always be used, in theory, to subtract any sources creating non-gaussianities.

The last possibility, and the only one that can accurately explain the substantial improvement of \glassviii over \glassvi, is that the GB sampler may have trouble converging, especially on the low-SNR but detectable sources in their highest density region.
The original, extremely flexible noise model of \glassvi could be to blame.
A very flexible noise model puts little pressure on the global fit to maintain the gaussianity of the residual by absorbing the non-gaussianity with low-SNR discrete sources.
It can instead locally increase the noise PSD, and since no instrument model or T-channel were involved, this overestimate is weakly penalized.
Another target of blame may be the GB amplitude prior. In trans-dimensional MCMC, the prior on SNR cannot extend to zero, or else a sea of poorly-constrained low-SNR sources will always be present and the sampler will struggle to converge.
However allowing more low-SNR sources may fit some of the remaining non-gaussianities.
Another large improvement in \glassviii are the set of proposals in the GB sampler, which include new one-to-many and many-to-one split and merge proposals.

So far in the global fit literature, goodness of fit metrics have focused on the resolvable sources at the posterior samples level and how well their catalogs match the source injections \cite{Littenberg:2023xpl,Katz:2024oqg}.
While these are good and important metrics, they are focused on the high-SNR sources.
It is one of the main messages of this paper that global fits should also be judged on the quality of their residuals.
To our knowledge, this effect has not yet been acknowledged in the literature.

Without mitigation measures, the presence of the non-gaussianity in the residuals means a stochastic sampler using the likelihood \eqref{eq:likelihood} is almost guaranteed a false detection, as it attempts to fit the incoherent excess power.
Even though this power is very unlikely to match any of the SGWB templates we study, some of the more peaked cosmological templates are fit at low-amplitude by this excess power, and it precludes setting accurate upper limits.

Several mitigation measures are possible: (1) shortening the time period of the STFTs removes more of the non-gaussianity due to non-stationarity, (2) a non-gaussian likelihood could be used (e.g. \cite{Sasli:2023mxr}), (3) modeling the non-stationarity-induced correlations in the likelihood, (4) improving the GB sampler so that it leaves fewer partially fit sources, or (5) vetoing the most non-gaussian $(t,f)$ data bins from the likelihood evaluation.

For this work we have taken approach (5), based on the same Anderson-Darling-based gaussianity measure in Figure \ref{fig:residual_gaussianity}.
 We test both the real and imaginary parts of the Fourier coefficients in small frequency windows in each time segment, and we veto bins that surpass a threshold probability. We take several values of this threshold to check the robustness of this procedure.
\begin{figure*}[t]
    \includegraphics[width=0.32\textwidth]{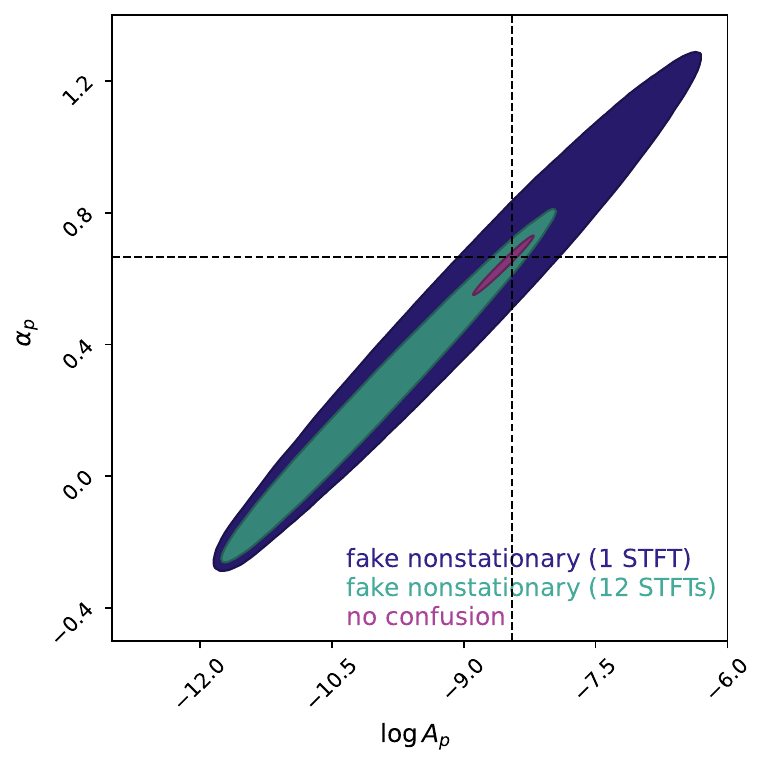}
     \includegraphics[width=0.32\textwidth]{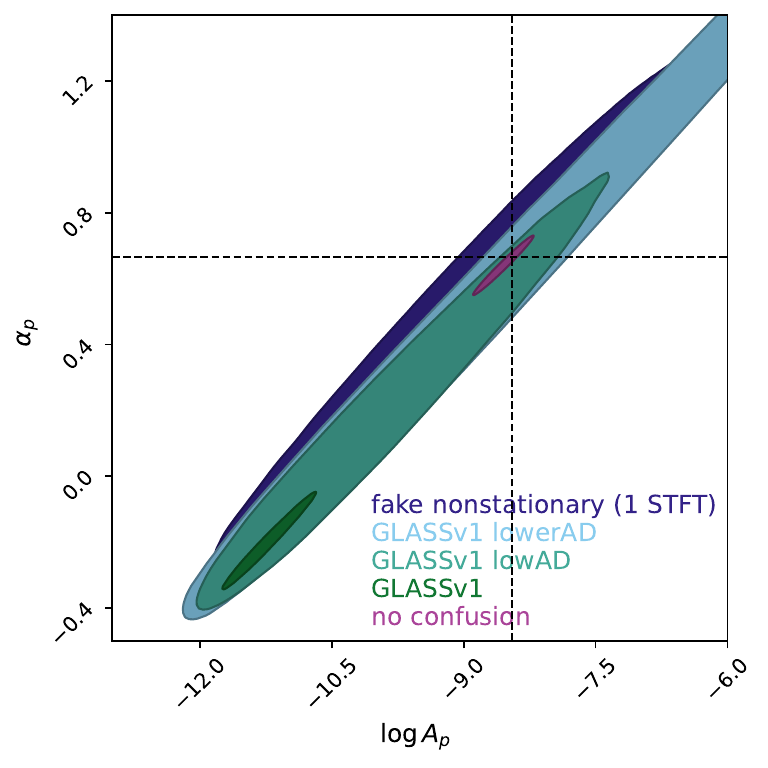}
     \includegraphics[width=0.32\textwidth]{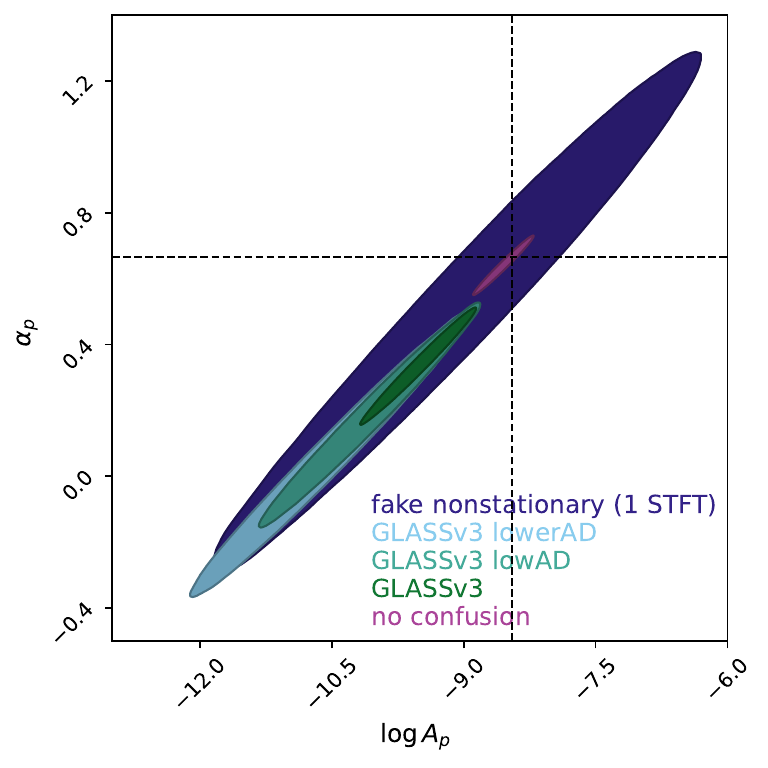}
    \caption{We show the attempted recovery of a powerlaw extragalactic background injection with the different galactic foregrounds considered in this work in a year of simulated data, marginalizing over all instrument and confusion noise parameters, and showing only $2\sigma$ contours for clarity. In the left panel, we look at synthetically generated confusion noise, while other panels show the two \glass  runs with different levels of gaussianity cuts applied to the data. For comparison, in all panels we also show the posteriors from simulated data with no confusion noise and an analysis assuming the synthetic confusion noise is stationary. Our injected powerlaw \eqref{eq:powerlaw} matches the fiducial parameters considered in the \radler data challenge LDC1-6, with $\log_{10} A_p = -8.45$, $\alpha_p = 2/3$.
    Mismodelling the confusion noise has a stark impact on our ability to recover the powerlaw SGWB: failing to account for the non-stationarity inflates the posterior, while non-gaussianity in the data biases the posterior away from the true value and generates false constraining power. Our technique for mitigating the non-gaussianity unbiases the posterior at the expense of reduced sensitivity. See the main text for details.
    }
    \label{fig:powerlaw_detection}
\end{figure*}
\subsection{Sampling}
\label{sec:pipeline-sampling}

For our proof-of-concept studies, we use the \texttt{Eryn} sampler \cite{Karnesis:2023ras}, which implements ensemble sampling and parallel tempering and is capable of efficiently sampling large dimensional parameter spaces -- ours is of dimension $N_s + \mathrm{nSTFTs} \times N_{\mathrm{ns}}$, where $N_s$ is the number of stationary parameters and $N_{\mathrm{ns}}$ is the number of non-stationary parameters, which in this work are considered independent in each STFT.
In the results we report in Section \ref{sec:results}, $N_{\mathrm{ns}} \leq 1$, although we did consider e.g., varying the confusion noise shape parameters, which drastically expands the dimension of the parameter space.

We generate a time-domain noise realization plus any stochastic signal injection with the algorithm described in Appendix \ref{sec:noise_generation}, and then add to these data the \texttt{GLASS} residual.

The data are then divided into short time segments and fast Fourier transformed, and in small windows of frequency in each time segment we compute the Anderson-Darling score. If the corresponding $p$-value surpasses a threshold, we skip over this bin in all likelihood evaluations.

We then take the user-supplied priors along with the likelihood \eqref{eq:likelihood_implementation}, evaluated as a product across the time segments, and begin generating posterior samples with \texttt{Eryn}. For the samples we present in this work, we take $\texttt{ntemps} = 10$ and $\texttt{nwalkers} = 6\times \texttt{ndim}$, a burn in of 10,000 steps, and 10,000 steps afterwards.
We then discard additional samples equal to three of the maximum autocorrelation length, and thin the resultant chains by half the minimum autocorrelation length. The minumum autocorrelation length is typically $\sim 2$ in the noise parameters, and the maximum is typically $\sim 30$ in the confusion noise amplitude in each STFT. On a 24 physical core 2020 Mac Pro, reaching this number of samples with 12 STFTs typically takes $\sim 4$ hours.
These sampling parameters are generally overkill, and our trace plots confirm that the measured low autocorrelation lengths accurately describe the statistical independence of the samples.

\section{Results}
\label{sec:results}

In this section we demonstrate our pipeline performance, showing the impact of the confusion noise on a selection of SGWB templates.

We first attempt a detection of a loud powerlaw signal of identical amplitude and spectral index to the one injected in the \radler data challenge LDC1-6 (from an estimate of the SOBHB background \cite{LISA:2022kgy}).
To see the effect of the different possibilities for the confusion noise, we inject each of them on top of the powerlaw SGWB and instrumental noise generated as in Appendix \ref{sec:noise_generation} and run the three data channels through our pipeline.

The posterior samples, marginalized over the instrumental noise and confusion noise parameters, are visible in Figure \ref{fig:powerlaw_detection}.
It is immediately apparent that, even with idealized instrument noise, accurate (or at least unbiased) modeling of the galactic foreground is crucial to be able to make any claim of a detection. 
After verifying that we can recover the background well in \radler-esque data with no galactic foreground, we inject both synthetic confusion noise (leftmost panel) and global fit residuals (other panels).
We see that the synthetic non-stationary confusion noise (generated as described in Section \ref{sec:data-generation})  substantially inflates the posterior but does not bias it, and that tracking the non-stationarity using our procedure with 12 STFTs improves the stochastic background recovery compared to a fully stationary analysis (1 STFT). Not shown, we scanned over $\mathrm{nSTFTs}=\{3,6,12,24\}$, and chose 12 as a nice middle ground without bias, and use 12 STFTs for all of our global fit residual analyses unless otherwise stated.

In the other panels we study the two \glass residuals.
In each panel, we perform a series of Anderson-Darling cuts on the data, as described in Section \ref{sec:pipeline-nongaussianity}, with three different values of $p_\textrm{crit}=\{1,10,15\} / N_\textrm{bins}$, the higher $p$-values corresponding to the ``lowAD'' and ``lowerAD'' labels respectively.
Reducing the non-gaussian contribution to the data helps unbias the estimated posterior, but also removes data from the likelihood and inflates the posterior when compared to the purely gaussian non-stationary case.
Also apparent from the \glass residuals is that the total amount of non-gaussian data cut decreases as we move to later stages in pipeline development.
\glassviii is mildly inconsistent with the injection even at the strictest gaussianity cut -- this may come from mismodelling the shape of the confusion noise or the fact that the piecewise-stationary analysis in this work does not accurately track the different $A$ and $E$ responses from an anisotropic sky distribution (cf. Figure \ref{fig:rae_comparison}).

In short, this case study makes apparent the effects of non-stationarity, non-gaussianity, and our mitigation measures on the posterior, as well as the possible limitations of a piecewise-stationary and isotropic assumption.

\begin{figure}[h]
    \centering
    \includegraphics[width=\columnwidth]{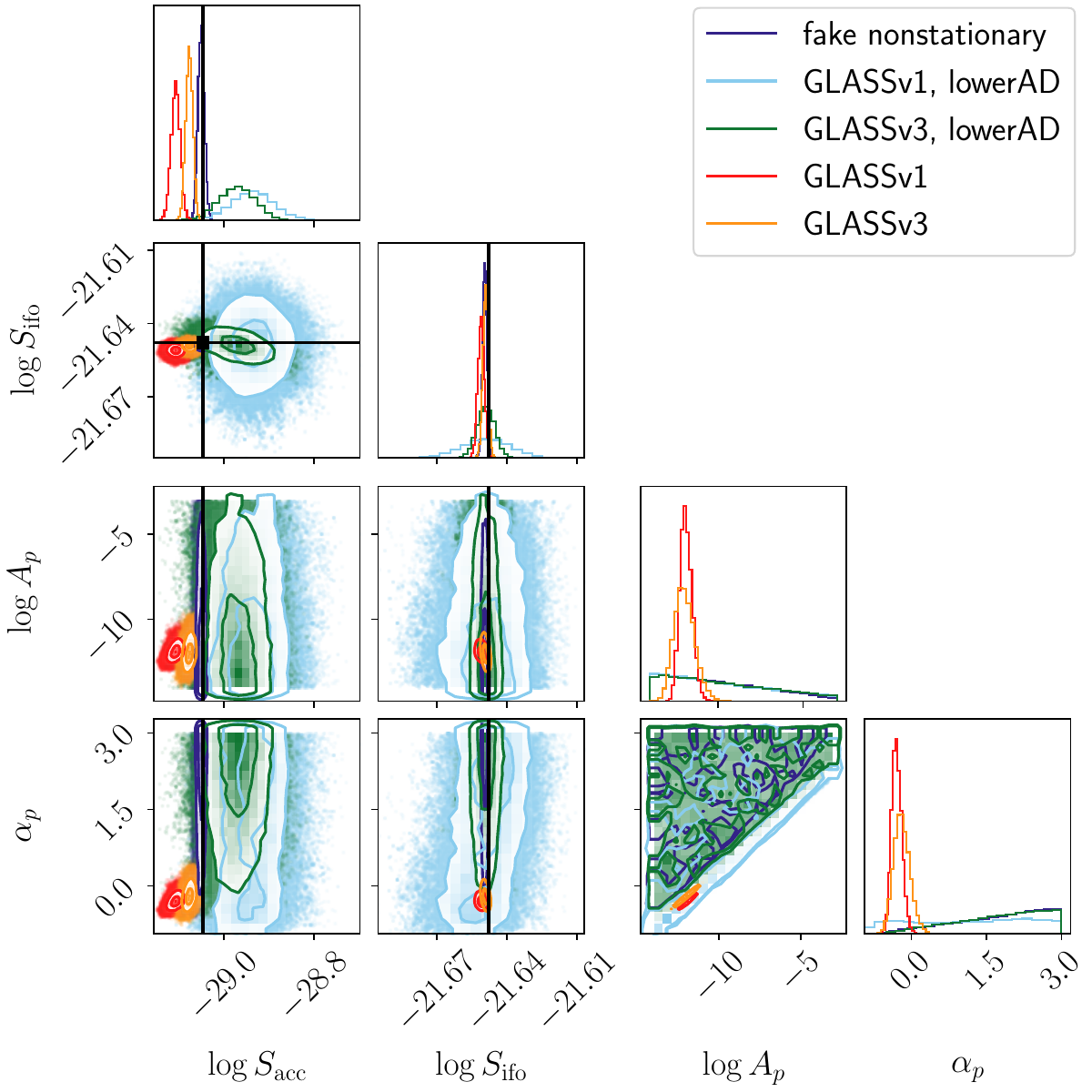}
    \caption{Powerlaw upper limits for several versions of the confusion noise drawn with $2\sigma$ contours, with injected parameters for the noise model drawn in black. We compare our aggressive non-gaussianity cuts (the ``lowerAD'' chains) to a permissive cut of the \glass residuals shown in red and orange. Allowing non-gaussianity in the data generates false detections at high significance, showing closed contours in the $A_p, \alpha_p$-plane. Our cuts on the data successfully mitigate the non-gaussianity well enough to allow us to set upper limits for a powerlaw SGWB.}
    \label{fig:powerlaw_sgwb_upper_limit}
\end{figure}

\begin{figure}[h]
    \centering
    \includegraphics[width=\columnwidth]{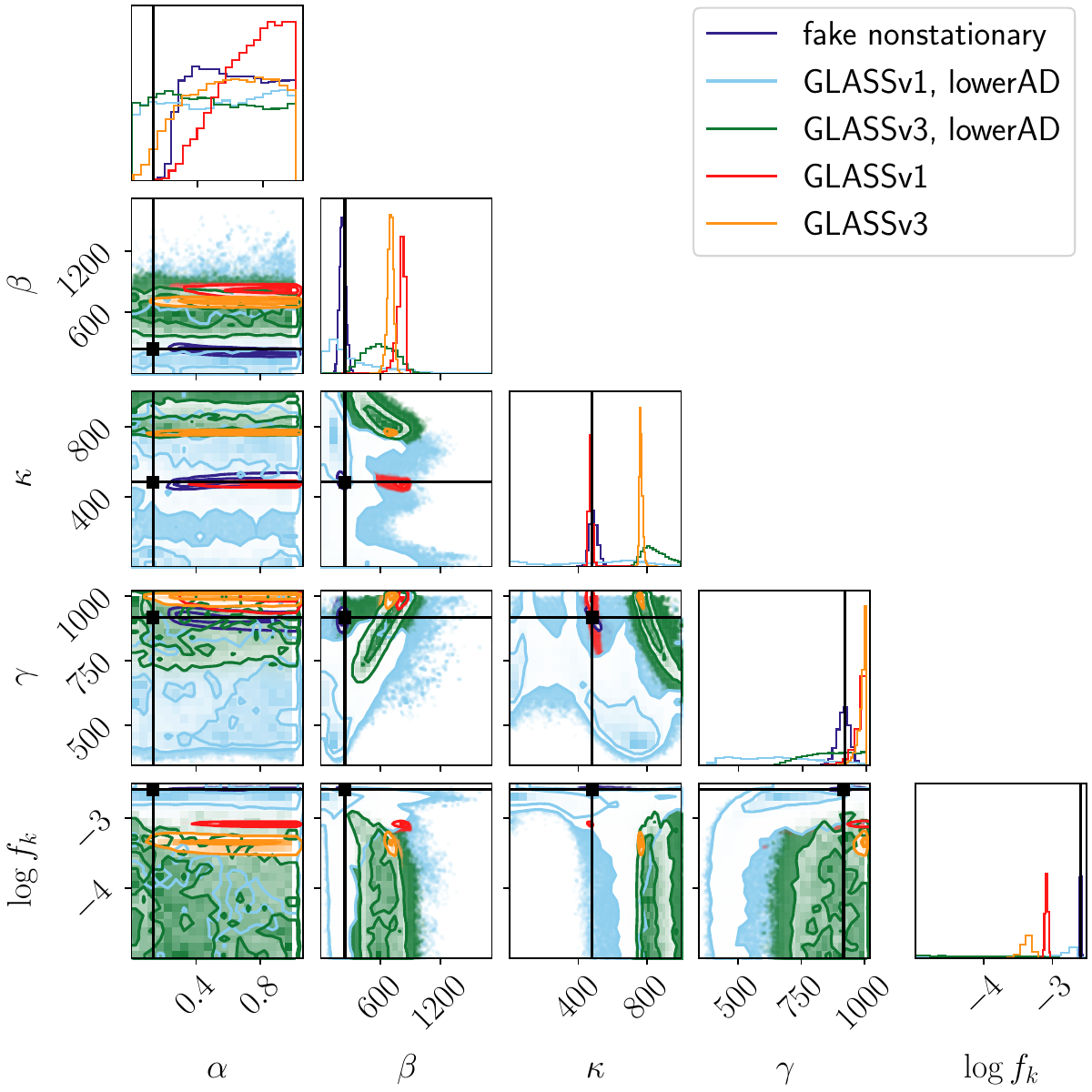}
    \caption{We show the confusion shape parameters during our powerlaw upper limit runs, the same chains as in Fig. \ref{fig:powerlaw_sgwb_upper_limit}. Injection values for the synthetic confusion noise are drawn in black. The global fit residuals with controlled non-gaussianity are not perfectly represented by the confusion template \eqref{eq:confusion}, with broad posterior support in most parameters, and a corresponding bias in the noise model parameters (cf. Figure \ref{fig:powerlaw_sgwb_upper_limit}).}
    \label{fig:powerlaw_sconf_upper_limit}
\end{figure}


\begin{figure}[h]
    \centering
    \includegraphics[width=\columnwidth]{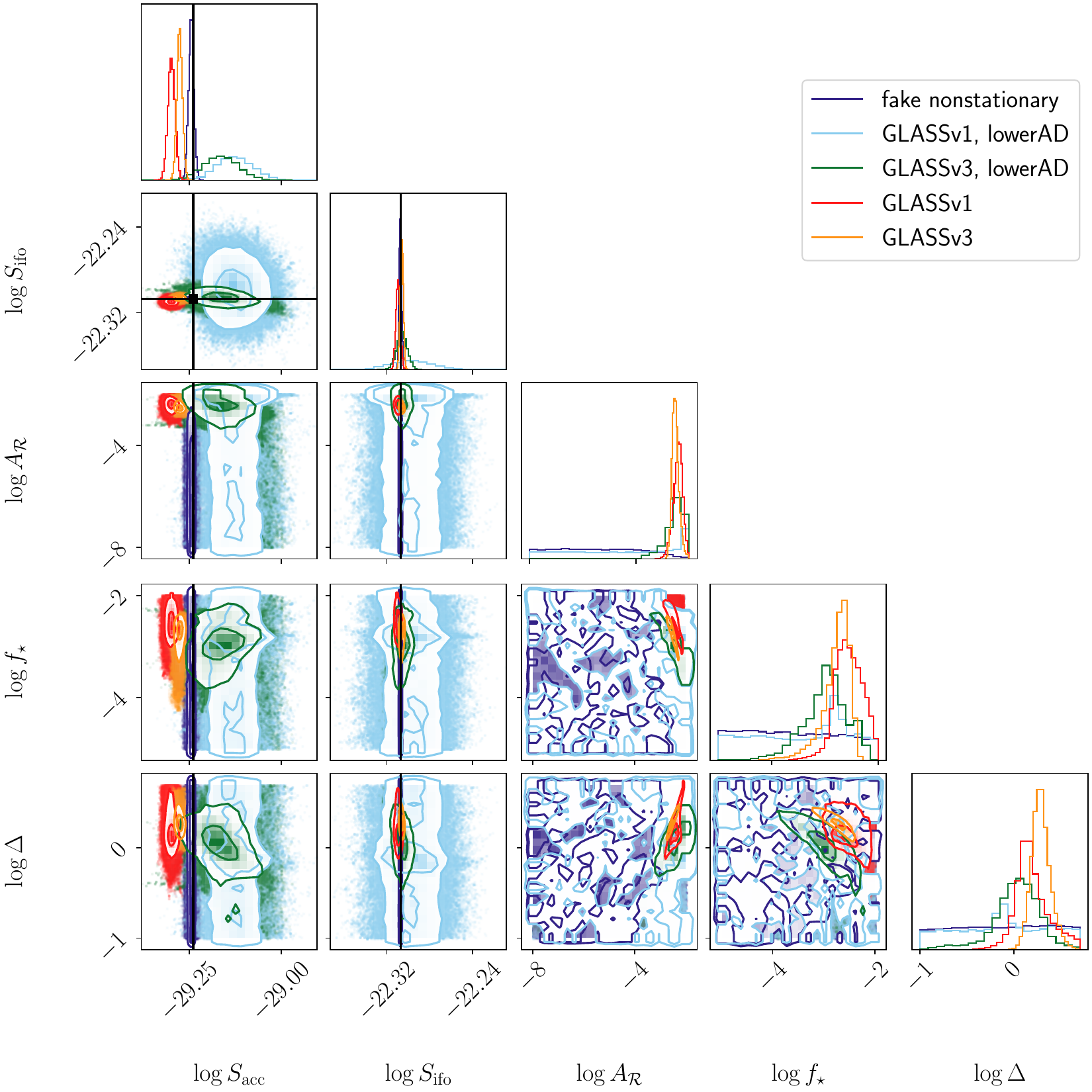}
    \caption{Lognormal scalar-induced SGWB upper limits for several versions of the confusion noise, with injected parameters for the synthetic non-stationary confusion noise drawn in black. We see that the disadvantage of large non-gaussianities in the data, as the \glass residuals have necessarily been cut and the posterior in $A_\mathcal{R},f_\star,\Delta$ begins to show a false detection. Not shown, the confusion shape parameters are similarly poorly constrained in the global fit residual cases as in Figure \ref{fig:powerlaw_sconf_upper_limit}}.
    \label{fig:lognormal_upper_limit}
\end{figure}
\begin{table}[h!]
    \centering
    \begin{tabular}{l|c}
         confusion noise & upper limit [$\log_{10}A_p$]  \\ \hline
         none & -10.41 \\
         synthetic non-stationary (1 STFT) & -10.05\\
         synthetic non-stationary (12 STFTs) & -9.76\\
         \glassvi, lower AD & -8.16\\
         \glassviii, lower AD & -9.26
    \end{tabular}
    \caption{Derived upper limits in simulated 12 month datasets for a powerlaw SGWB with the form \eqref{eq:powerlaw} and various assumptions for the confusion noise. The quoted limits are 95-percentile upper bounds, coming from the posterior of the final STFT fit to a kernel density estimator and conditioned on $\alpha_p=2/3$. }
    \label{tab:powerlaw_upper_limits}
\end{table}

\begin{table}[h!]
    \centering
    \begin{tabular}{l|c}
         confusion noise & upper limit $[\log_{10} A_\mathcal{R}]$ \\ \hline
         none & -3.40 \\
         synthetic non-stationary, 1 STFT & -3.32\\
         synthetic non-stationary, 12 STFTs & -3.00\\
         \glassvi, lower AD & -2.26$^*$ \\
         \glassviii, lower AD &-2.10$^*$ \\
    \end{tabular}
    \caption{Derived upper limits for a scalar-induced SGWB from a lognormal peak. We quote 95\% upper limits in $A_\mathcal{R}$, fitting the posterior to a kernel density estimator and conditioning on $f_\star=0.2\, \mathrm{mHz}$ and $\Delta=0.3$. Unfortunately, even our most aggressive gaussianity cuts were not sufficient to be able to confidently set upper limits using the \texttt{GLASS} residuals -- the posteriors are peaked at false detections. These may come from a mismatch in the confusion noise shape, even after skipping over the most non-gaussian data.}
    \label{tab:lognormal_upper_limits}
\end{table}

Next we look at upper limits, since they are likely to be the most challenging data analysis task for an SGWB search, and are the most important data product for the community absent a detection.

In Figures \ref{fig:powerlaw_sgwb_upper_limit}, \ref{fig:powerlaw_sconf_upper_limit} and Table \ref{tab:powerlaw_upper_limits} we display the upper limits for a powerlaw SGWB for the various confusion noises, comparing the synthetic confusion noise (with injection values marked on the figures) to the \glass residuals.
In our most aggressive gaussianity cuts, we are able to successfully set upper limits using the global fit residuals, at a comparable amplitude to the synthetic confusion noise case. In Table \ref{tab:powerlaw_upper_limits}, we quote 95\% confidence limits on the powerlaw's amplitude fixing $\alpha_p=2/3$.
We also show a worst case scenario in the figures in red and orange: a false detection of a powerlaw SGWB (!) from using either of the \glass residuals with a too-permissive gaussianity cut.
We also look at the confusion shape parameters, noting that they are much worse constrained in the \glass confusion noises than in the synthetic case.  \glassviii used the same analytic model for the confusion noise (with $\beta\equiv0$), so this mismatch is surprising to us and indicates the confusion noise shape may not be so easily described as in \eqref{eq:confusion}.

Unfortunately, this performance is not matched on the other SGWB shapes we consider. A powerlaw is a broadband signal and insensitive to local features in the residuals, which can generate false detections for more narrow-band signals.
In Figure \ref{fig:lognormal_upper_limit} and Table \ref{tab:lognormal_upper_limits}, we see the equivalent upper limits attempted for the scalar-induced SGWB from a lognormal overdensity.
Because this is a narrow-band background, it is much more sensitive to misfits of the confusion noise or other residual signal power in the data. Even in our most aggressive gaussianity cuts, the posterior shows a false detection in the lognormal SGWB parameters in both of the \glass confusion noises.
The posteriors peak around $\sim 3$ mHz, suggesting that there is excess power in the data there even after our gaussianity cuts.

\begin{figure}[h]
    \centering
    \includegraphics[width=\columnwidth]{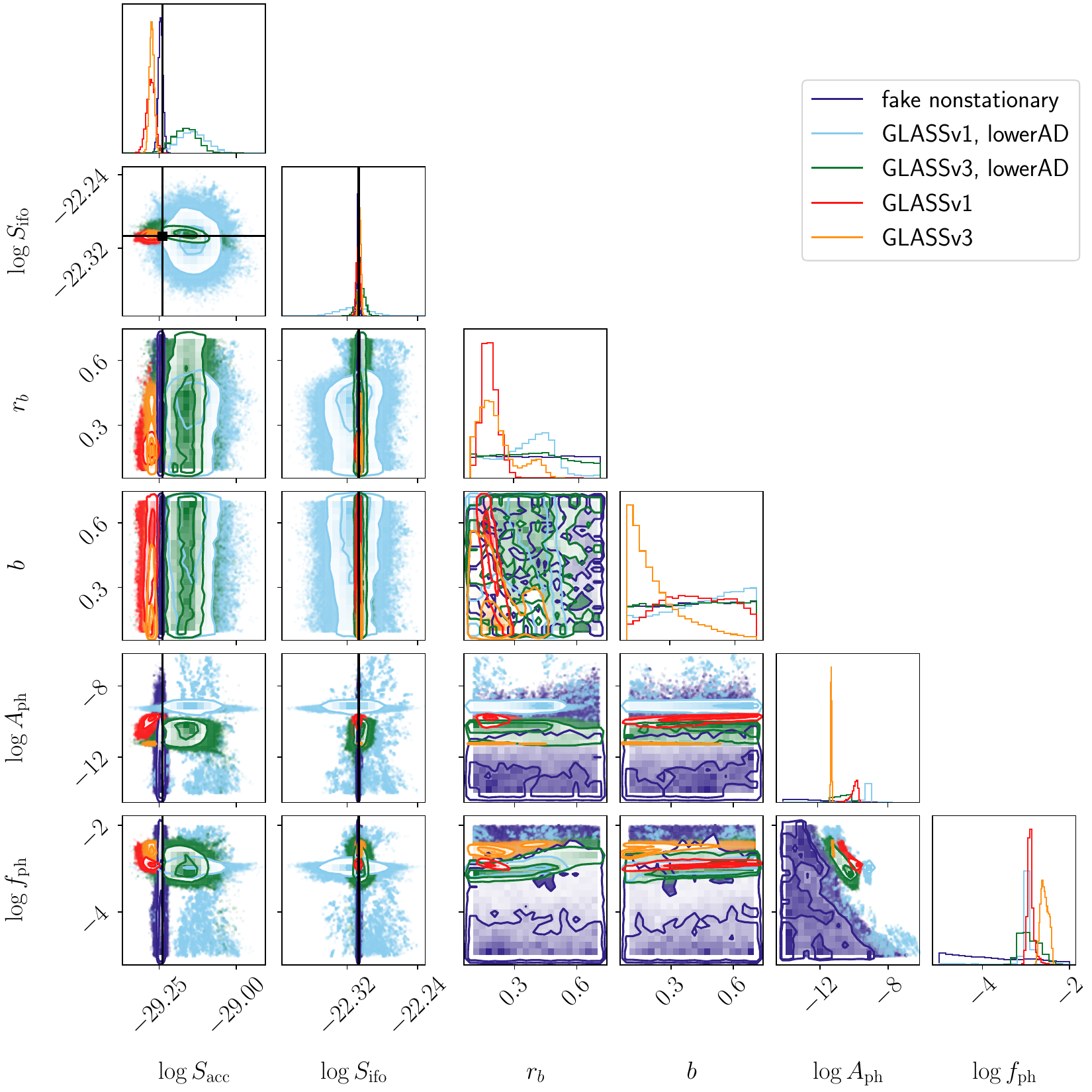}
    \caption{Attempted upper limits for the phase transition template under different confusion noises. We see a false detection even in our most conservative gaussianity cuts, indicating to us that un-modeled excess power is still in the data. See main text for discussion.}
    \label{fig:pt_upper_limit}
\end{figure}
\begin{table}[h]
    \centering
    \begin{tabular}{l|c}
         confusion noise & upper limit [$\log_{10}A_\mathrm{ph}$]  \\ \hline
         none & -11.46 \\
         synthetic non-stationary (1 STFT) & -10.65$^*$\\
         synthetic non-stationary (12 STFTs) & -11.36\\
         \glassvi, lower AD & -11.53$^*$\\
         \glassviii, lower AD & -12.21$^*$
    \end{tabular}
    \caption{Derived upper limits in simulated 12 month datasets for a phase transition SGWB with the form \eqref{eq:pt} and various assumptions for the confusion noise. The quoted limits are 95-percentile upper bounds on the amplitude, coming from the posterior fit to a kernel density estimator and conditioned on $r_b=0.15, b=0.65, f_\mathrm{ph}=0.2\, \mathrm{mHz}$. In all of the global fit residuals and in the stationary approximation of the synthetic confusion noise, the posterior shows false detections and cannot set true upper limits.}
    \label{tab:pt_upper_limits}
\end{table}

As a middle ground, we also consider a middle bandwidth SGWB.
In Figure \ref{fig:pt_upper_limit} and table \ref{tab:pt_upper_limits}, we repeat the same analysis for the first-order phase transition template.
Again, the more strongly peaked background is more prone to false detections, by some un-modeled power in the data at $\sim 3\, \mathrm{mHz}$.
Additionally, we see a false detection when treating the synthetic confusion noise as a stationary source, presumably due to the induced non-gaussianity from non-stationarity.

These false detections make it absolutely clear to us that these global fit residuals are not yet ready for the extreme scrutiny provided by an upper limit search.
We also caution that narrow-band SGWBs are much more demanding of the residuals than the more commonly studied powerlaws in the data analysis literature.

\begin{figure*}[t]
    \centering
    \includegraphics[width=0.49\textwidth]{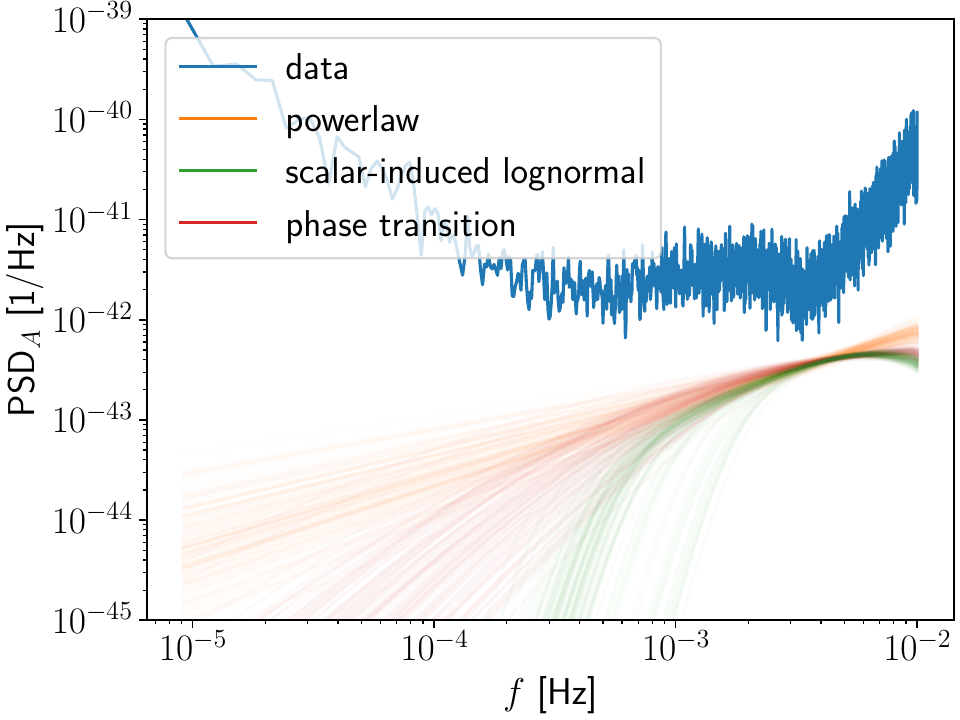}
    \includegraphics[width=0.49\textwidth]{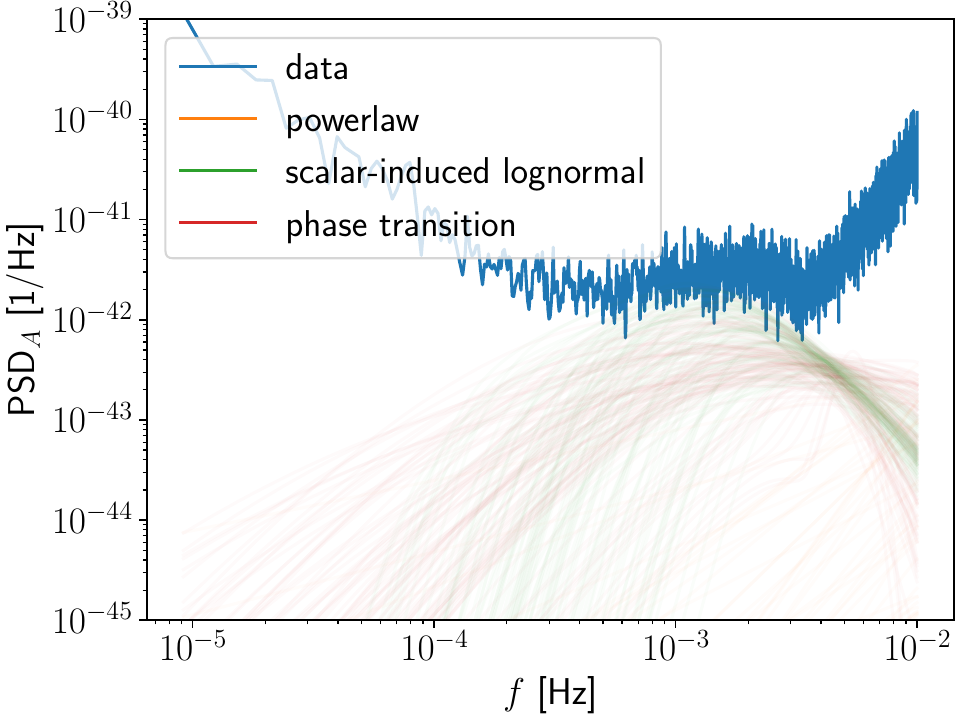}
    \caption{\textbf{Where are the false detections coming from?} We show the A-channel PSD of the \glassviii data along with the estimated SGWB from 100 fair draws of parameters from the posterior of each of the upper limit runs. On the left we perform a minimal gaussianity cut while on the right panel we performed our strictest cut (\texttt{lowerAD}). 
    In the lax cut, in all cases, the false detections fit an excess of power around $\sim 4$ mHz. In the strict cut, the situation is markedly improved, with the powerlaw SGWB no longer finding excess power in the data. Unfortunately the more narrow-band SGWBs still pick up excess power.
    Again, the region of the confusion noise around $\sim 4$ mHz is responsible for the false detections, although they are of lower significance.
    }
    \label{fig:all_false_detections}
\end{figure*}

The false detections can be summarized in Figure \ref{fig:all_false_detections}, where we plot the last month of \glassviii data along with 100 fair draws from each upper limit attempt's posterior.
On the left we show the most permissive gaussianity cut, which has false detections in all three SGWB types. On the right, we show the strictest gaussianity cut which still has false detections for the more narrow-band SGWBs.
In both cases the false detections show localized excess power around the $4$ mHz region, with the \texttt{lowerAD} runs indicating a lower significance of false detection.
This reduced significance is also visible in the corner plots, Figures \ref{fig:lognormal_upper_limit} and \ref{fig:pt_upper_limit}. 

\section{Conclusions and Future Work}
\label{sec:conclusions}

In this work we set out to study how to perform stochastic background searches for LISA.
Due to the global nature of the data analysis problem, the literature has often proposed hierarchical searches working with the global fit residuals. We discuss many of the challenges of these searches and the nature of the residual in Section \ref{sec:residual}.
We built a pipeline to perform SGWB parameter estimation, which we describe in Section \ref{sec:implementation}, and verified that we can successfully inject and recover SGWBs using a synthetic confusion noise in Figure \ref{fig:lognormal_fit_violin}.
We also include mitigation measures for the non-gaussianity in the data (visible in Figure \ref{fig:residual_gaussianity}) and model the non-stationarity of the data (visible in Figure \ref{fig:rae_comparison}) by breaking it into short time segments.

To test how well this procedure would work for the global fit, in Section \ref{sec:results} we consider the synthetic confusion noise as well as two residuals from \glass. We use the one from the results presented in \cite{Littenberg:2023xpl} (\glassvi), and a later run after significant pipeline changes and development (\glassviii).
As seen in Figure \ref{fig:powerlaw_detection}, we can recover a loud detection (an estimate of the SOBHB background in \radler) in every case of the confusion noise, but with a significant bias in the parameters of the \glass recoveries unless we mitigate the non-gaussianity in the residual.
Upper limits are more challenging -- we are able to set valid upper limits for a powerlaw SGWB in Figure \ref{fig:powerlaw_sgwb_upper_limit}, but despite our best efforts we see false detections of more narrow-band SGWB shapes in Figures \ref{fig:lognormal_upper_limit} and \ref{fig:pt_upper_limit}.
All of the false detections are summarized in Figure \ref{fig:all_false_detections}, where it is apparent that the most non-gaussian region of the data around $4$ mHz is responsible.

We conclude that the dream of a hierarchical SGWB analysis is a very challenging one, sensitive to any non-gaussianities or excess power in the global fit residuals. This is especially true for the narrow-band SGWBs of interest to many cosmologists.
The shape and gaussianity of the real confusion noise do not match our expectations (cf. Eq. \eqref{eq:confusion}), and can generate false detections, some at high significance.
From a data analysis perspective, the residual is a very noisy place, containing at a minimum: bursts, glitch and gap residuals, any waveform inaccuracies or beyond GR effects, any mismodeling of the instrument, and any poorly subtracted sources from the global fit.
In using the \sangria data, we know that our waveforms are accurate, there are no glitches or gaps, and we know the noise model. Yet we still see false detections of narrow-band SGWBs. An SGWB search is a very high level of scrutiny to put on a residual.

We now comment on some future directions:
\subsection{Global fit integration and online search}
The pipeline we present here has the capability to replace the current \glass noise model, and the current Python-language bindings around the C-language likelihood and templates could be readily swapped for the equivalent \glass samplers.
The latest refactoring of \glass supports $X$, $Y$, $Z$ channels and an arbitrary covariance matrix, so a rotation of \eqref{eq:likelihood} is readily applicable.

Such an online search would provide the \glass GB sampler the best chance to identify excess power in the current residual and attempt to fit it with templates.
Similarly, our current hierarchical analysis works with only one fair draw of the \glass posterior, but an online search would naturally pass the SGWB sampler many realizations of the residual and properly marginalize over the different possible source contents of the data.

It would also be possible to build reversible jump proposals to reduce the possibility of false SGWB detections by, e.g., swapping an SGWB for a collection of low-amplitude GBs or targeting GB proposals at areas of the residual with high non-gaussianity.

An online search is not guaranteed to have better results than the prototype hierarchical one we present here, but having the SGWB and GB samplers exchange information is likely to improve convergence and hopefully alleviate some of the false detections.

\subsection{Confusion noise modeling}

The current confusion noise model \eqref{eq:confusion} is an empirical fit to an iterative source subtraction procedure \cite{Timpano:2005gm}.
There is no reason, \textit{a priori}, why this fit should perfectly describe the shape of the confusion noise left after the quite different procedure of trans-dimensional MCMC in \glass, and it perhaps should not be not surprising that the shape of \glass's confusion noise could differ.
We also know that there will be other populations in the confusion noise \cite{Scaringi:2023xpm}, and very likely we will not have modeled every source type to be able to confidently predict its shape.

These two considerations lead us to the conclusion that an ideal confusion noise model should have a high degree of flexibility, to fit the likely present unknown populations and allow the global fit pipeline to fit as many sources as possible.
It may be possible to use the resolvable sources to learn some population-level information and put a strong prior on the shape of the confusion noise of the unresolved population.

In this work we have modelled the confusion noise as being an isotropic intrinsically non-stationary source while in reality it is an extrinsically non-stationary but intrinsically stationary anisotropic source. The A and E channel amplitudes are equal in the isotropic case while an anisotropic signal has differing responses $R_A(t) \neq R_E(t)$ (cf. the expected amplitudes in Figure \ref{fig:lognormal_fit_violin}). We see small biases in our detection case study in Figure \ref{fig:powerlaw_detection}, which may be due to this effect.
Work is in progress to model the galactic foreground as a true anisotropic source inside of \glass.
Slotting in this improvement to our SGWB pipeline could be done with a small adjustment of the likelihood and response functions.

These improvements to the confusion noise modelling are likely \textit{requirements} to be able to meet LISA's stochastic background search objectives and be able to set upper limits.

\subsection{Instrumental modeling}
The LISA instrument is likely to experience interruptions in its ability to synthesize the constellation, from routine antenna-repositioning, micrometeorite impacts, or other effects \cite{redbook}.
After some of these gaps, some of the instrument noise parameters may shift.
There may also be other sources of non-stationarity in the instrumental noise on longer timescales.
One conservative option for handling these without bias is to chunk the data in time and treat the noise model as piece-wise stationary with independent parameters in each short time segment.
This is exactly the same procedure we've used to track the non-stationarity of the galactic foreground and is already supported in our current pipeline.

The noise knowledge of the instrument may not be perfectly described by an analytic function as we have used here, but only known and characterized approximately.
In that very likely scenario, noise knowledge uncertainties become a crucial part of stochastic background detection, and our pipeline should implement an ability to estimate the noise model simultaneously. A proof-of-concept implementation is in \cite{Baghi:2023qnq}.
Losing noise model confidence will make SGWB recovery significantly more challenging, but the different TDI responses of signal and noise allow for inference nonetheless. At low frequencies the noise and signal TDI transfer functions become degenerate, but they become more orthogonal at high frequency.
\subsection{Model selection}

One other interesting question would be determining the source of any stochastic signal in the data.
In a stochastic template search like the one we have here, this would naturally come out of having these signals in a reversible-jump MCMC scheme, as templates are continuously tried against the current residual.

\subsection{Understanding and exploiting the non-gaussianities}
The Milky Way's population of galactic binaries generates both the resolvable and confused sources. If we are able to better understand under what conditions a source is confused, we may be able to use the observed GBs and a galactic population model to set a strong prior on the confusion noise shape and statistics of its sources.
If there are additionally non-gaussianities in the confusion noise, then it contains some extra information about the sources in the highest frequency bins and may further constrain the unresolved population.
Also, a joint resolvable and unresolvable GB population model may help avoid any bias from the practical necessity of disallowing priors that include low-SNR sources.

Another effect that may be a signature of non-gaussianities is a higher bias of the discrete sources above non-guassian regions of the noise. This bias may preclude proper parameter estimation of these sources if the non-gaussianities aren't taken into account.
If this bias is not taken into account and is visible in the final catalog of sources, it may hint to non-gaussianities in the biased regions and may allow recovering additional information about the unresolvable population.

\begin{acknowledgments}
We would like to thank Michael Katz, John Baker, Ira Thorpe, Riccardo Buscicchio, Nikolaos Karneisis, Olaf Hartwig, Jonathan Gair, Mauro Pieroni, Tiina Minkkinen, and Alexander Criswell for helpful advice and discussions during the course of this work. RR was supported by an appointment to the NASA Postdoctoral Program at the NASA Marshall Space Flight Center, administered by Oak Ridge Associated Universities under contract with NASA.
\end{acknowledgments}

\appendix

\section{Posterior propagation}
\label{sec:GMMs}
In an earlier version of this pipeline, we did not use the product likelihood \eqref{eq:likelihood_implementation}, but instead repeated sampling in each time segment, encoding information from the previous time segments in the prior.
Some version of this procedure will likely be necessary in a production-ready LİSA global fit, as repeating the full analysis each time data is downlinked is a large and hopefully avoidable computational burden. Indeed, \glass uses a sort of time-annealing based proposals based on shorter time segments, and \erebor also uses similar proposal-based schemes for propagating preparatory analyses forward to the full data.

It is straightforward to show that using the posterior of a previous time segment as the prior of the next time segment should result in the same final posterior.
Consider a case with two STFTs, where the data vector is split in half $d = \{d_1,d_2\}$. Then we can split Bayes' Theorem into two iterations: $p(\theta|d) = p(d | \theta) p(\theta) / p(d) = p(d_2 | \theta) p(\theta|d_1)/p(d_2) = p(d_2 | \theta)
p(d_1 | \theta) p(\theta) / p(d_1) p(d_2)$.
Here the final posterior $p(\theta|d)$ has been built out of these two intermediate sampling steps, but we can similarly extend this to any number of intermediate sampling steps.
When done this way, we only propagated the posterior forward for the stationary parameters and considered the nonstationary parameters to have independent priors in each time segment. 

This method drastically reduces the dimension of the parameter space, instead of $N_{\rm STFTs} \times N_{\rm ns} + N_{\rm s}$ parameters in the sampler, we have only $N_{\rm s} + N_{\rm ns}$, but repeat sampling $N_{\rm STFTs}$ times.
This procedure allows us to capture some non-stationary features with a small modification of a purely stationary analysis.
İn practice, this is usually a net gain in speed, as there are many fewer uncorrelated directions in parameter space (e.g. the nonstationary parameters in STFT 1 vs those in STFT 8 by default are sampled simultaneously in the product likelihood method but can be sampled independently in the posterior-propagating method).

\begin{figure}[h]
\centering
\includegraphics[width=0.9\columnwidth]{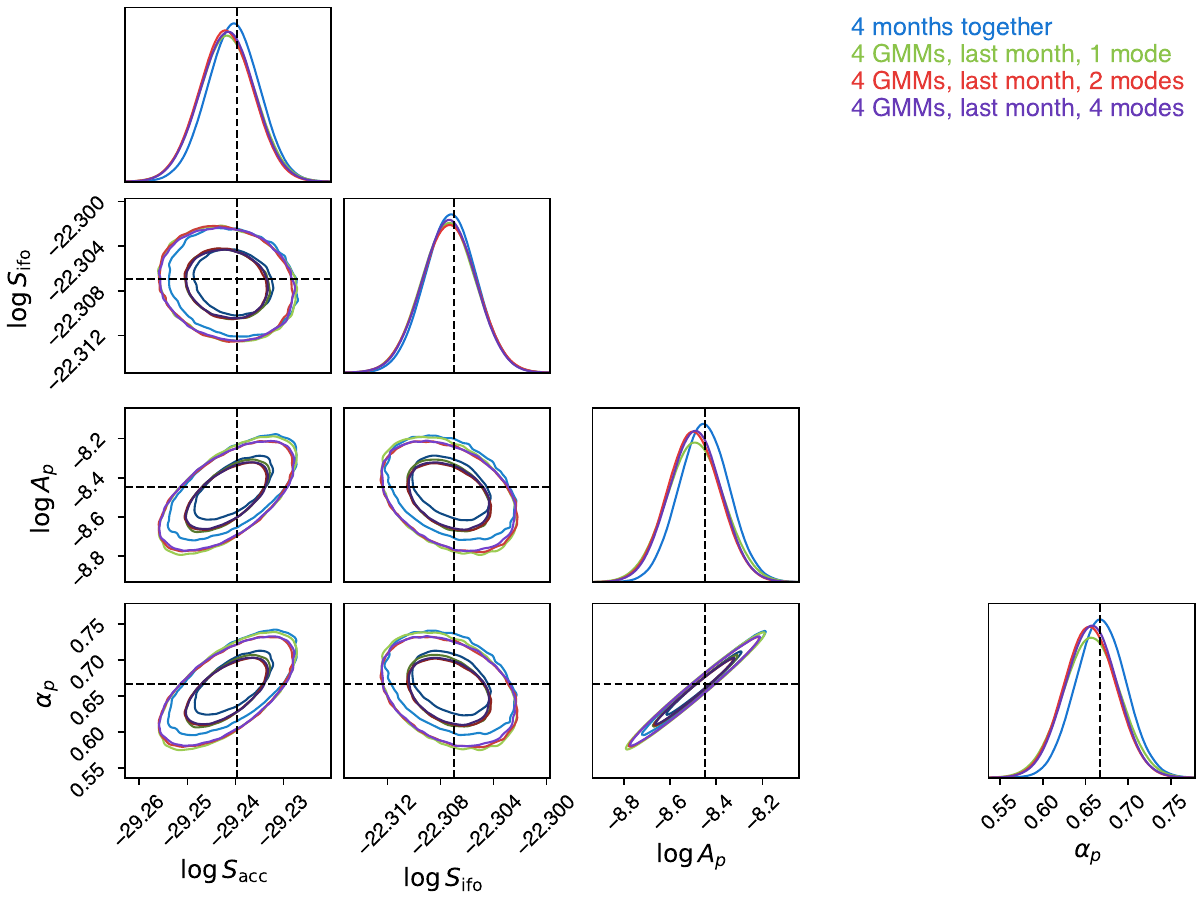}
\caption{A comparison of 4 months of stationary synthesized data with a powerlaw injection, analyzed with and without the STFT and GMM procedure we describe.}
\label{fig:GMM_comparison}
\end{figure}

Of course, posterior propagation can only work well when the posterior is reproduced accurately.
In practice, MCMC and other stochastic sampling algorithms do not estimate the true posterior as a function, but only sample it. Expressing the prior of the next STFT as a function then requires some kind of fitting procedure to generate new samples with the same distribution.
We chose to use gaussian mixture models (GMMs), since our posteriors are often close to gaussian and the implemenations are readily accessible in \texttt{scipy} \cite{scipy}. We use the \texttt{BayesianGausssianMixture} class because it fits for the individual gaussian components' weights in addition to their means and covariance matrices. We allow up to $N_s$ (the number of stationary parameters) gaussian components.
Figure \ref{fig:GMM_comparison} shows a test of stationary data modelling instrumental noise and a powerlaw cosmological background generated according to Appendix \ref{sec:noise_generation}. In the figure, we compare four months of data, either analyzed directly in one chunk, or split into four STFTs using our posterior propagation procedure. The estimated posterior distributions compare very favorably, but this example is nearly ideal with a very gaussian posterior.

\begin{figure}[h]
\centering
\includegraphics[width=0.9\columnwidth]{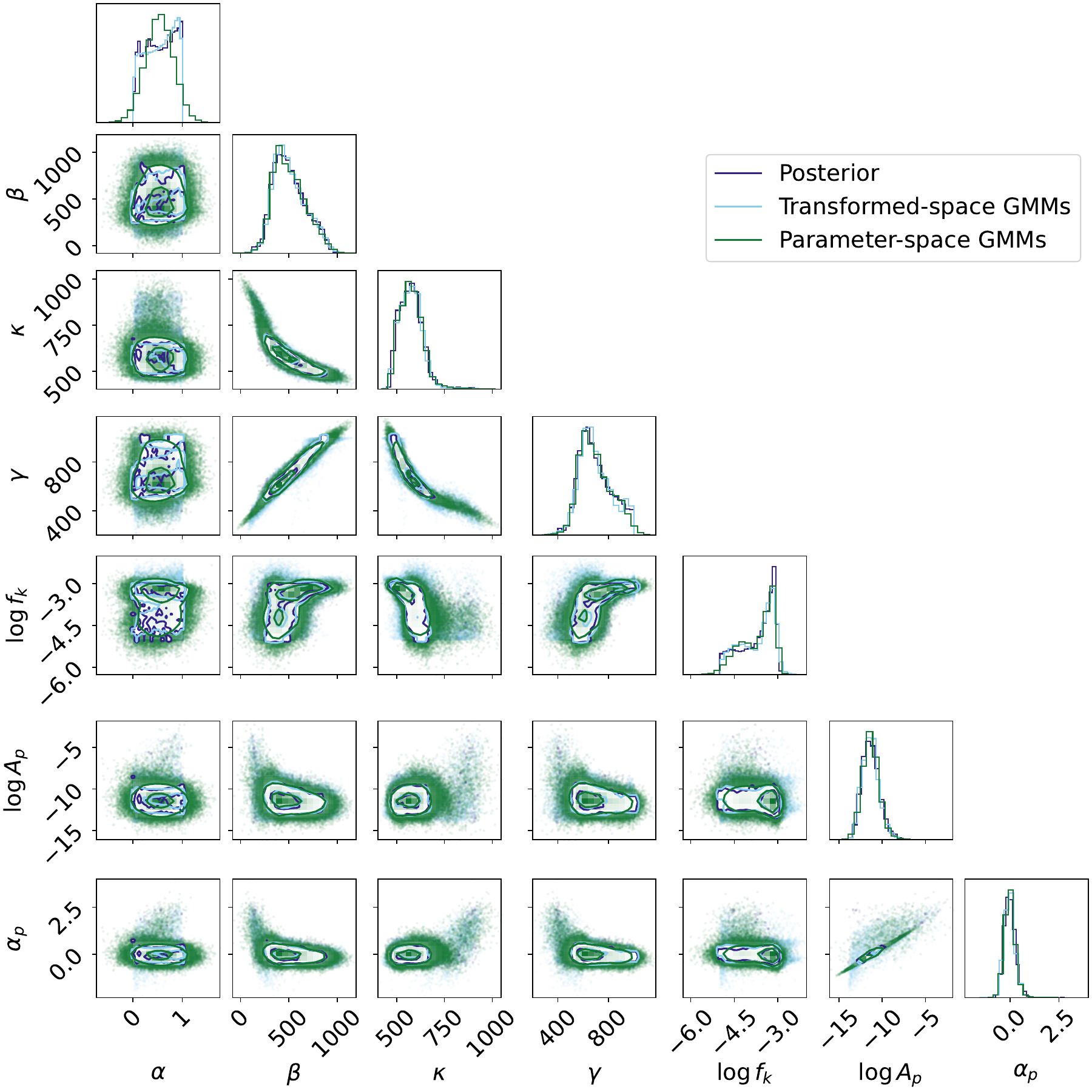}
\caption{A comparison of a posterior fitting attempt with a real space GMM and a logistically-mapped GMM. This is an upper limit attempt on a \glass residual, and therefore has a fairly non-gaussian posterior. While the logistic mapping is an improvement, the fidelity of either fit is, in our opinion, not robust enough to set upper limits.}
\label{fig:logisticGMM}
\end{figure}

For more non-gaussian posteriors, we also attempted a logistically mapped GMM fit.
The logistic mapping transforms the posterior samples into another coordinate system in which they are hopefully more gaussian, and then samples the prior in this new space, transforming the samples back before passing them to the sampler. 
We used the mapping $\theta_i = \frac{P_i}{1+\exp(-\theta_i^\prime)} + Q_i$, where $\theta_i$ are the parameters in our original coordinates, $\theta_i^\prime$ are the transformed parameters, and $P_i$ is the range of the priors and $Q_i$ is the minimum allowed value in the prior in coordinate $i$. One complication of this procedure is that the prior probability evaluated in the transformed space needs a Jacobian to account for the coordinate transformation. So the prior value in the original coordinates $p(\theta_i) = |\frac{\partial \theta_j^\prime}{\partial \theta_k} | \times p(\theta^\prime_i)$.
The performance of these logistically mapped GMMs can be compared to that of real-space GMMs and the raw posterior samples in a more challenging case in Figure \ref{fig:logisticGMM}.

While a substantial improvement, we considered the posterior reproduction fidelity insufficient for this paper, and instead chose to proceed with our full analysis with the product likelihood.

Neural flow posterior estimation and other methods are also popular in the literature, and we would consider them if there were plug-and-play packages available. We also note that neural flow posterior estimation is unlikely to be perfect (e.g. the problem of filaments between modes of the posterior is well-documented).

Unless the situation markedly improves, it is our opinion that the production global fit should use these posterior estimation techniques either as proposals or as priors in rapid-response analyses, but not as priors in any quoted full-confidence data products.

\section{Data generation}
\label{sec:noise_generation}

In this work we often generated a colored noise or stochastic signal from a given power spectral density, and we describe our algorithm and conventions for doing so here.

If we have a timeseries with samples $x_{ij}$, where $i$ labels the channel and $j=1\ldots N$ is the sample index, we can define the power spectral density as $\mathrm{PSD}_{ik} \equiv \frac{2 \mathrm{dt}}{N} |X_{ik}|^2$, where we define the discrete Fourier transform with no normalization in the forward direction $X_{ik} = \sum_{j=0}^N x_{ij} e^{2\pi i k j/N}$ and $x_{ij} = \frac{1}{N}\sum_{k=0}^N X_{ik} e^{-2\pi i k j/N}$,  where $X_{ik}$ are the Fourier coefficients and there are $N$ samples of the signal in time with uniform time delta $\mathrm{dt}$. This is the default transform in the \texttt{numpy} package \cite{numpy}.

For a stationary gaussian-random signal or noise with an expected PSD with coefficients $S_{ik}$, the real and imaginary parts of each $X_{ik}$ must both be gaussian-distributed, with standard deviation given by the square root of half of the PSD.
In other words, 
\begin{align}
X_{ik} \sim \sqrt{\frac{S_{ik}}{2}}\left(N(0,1) + i N(0,1)\right),
\label{eq:random_FFT_from_PSD}
\end{align}
where $N(0,1)$ is an independent unit normally distributed random variable \footnote{Alternatively, the $X_{ik}$ can be viewed as complex random variables with uniform-random phases and Rayleigh-distributed amplitudes \cite{Buscicchio:2024wwm}} and $S_{ik}$ is the theoretically desirable average PSD over many realizations. These $X_{ik}$ will match the likelihood \eqref{eq:likelihood} with $S_{ik}$ as their variance.

In practice we generate our stationary signals and noise first in the frequency domain via \eqref{eq:random_FFT_from_PSD}, and then use these Fourier coefficients and an inverse transform to generate a timeseries.
In order to ensure that our data have mean zero, we enforce that the DC frequency bin is zero, and then apply \texttt{numpy.fft.irfft} on the coefficients $X_{ik}$ to obtain our strain timeseries.
The periodogrammed PSD of this timeseries (with a boxcar window) will match our realization of it constructed earlier to numerical precision.

When generating multiple signals, e.g. a powerlaw SGWB injection plus instrumental noise, we generate each of the corresponding timeseries independently and add them. Similarly, the \texttt{GLASS} residuals are output as strain timeseries and may be directly added to the data generated here.
When we generate nonstationary confusion noise, we first generate completely stationary confusion noise following the procedure here, then multiply the end strain timeseries by $\sqrt{r_A(t)}, \sqrt{r_E(t)}$ from \cite{Digman:2022jmp} in the $A$ and $E$ channel respectively.

Once the final data timeseries has been generated, it is chunked into time segments and periodogrammed following the description in Section \ref{sec:implementation}.
Our periodogramming severely reduces the computational cost of evaluating the likelihood by assuming the data is purely stationary within one time segment and averaging the PSDs of several smaller sub-segments. We use the implementation in \texttt{scipy.signal.welch} \cite{scipy,1161901}, with typically $2^{16}$ time samples per segment and a total number of segments of $N_\mathrm{segs} \sim 8$ for month-long STFTs and the \sangria sample rate of $5$ seconds.
In principle, the bins in the Welch method are averages of squared gaussian variables and should be $\chi^2$-distributed, but with sufficiently many averages these become gaussian with an extra scaling prefactor of $N_{\mathrm{segs}}$ \cite{Pieroni:2020rob}, as we have in \eqref{eq:likelihood_implementation}.
We typically use a Hann window in each sub-segment, which nearly minimizes spectral leakage at the cost of somewhat reducing the data's constraining power \cite{numericalrecipes}.
It is possible to overlap the small time sub-segments in the Welch method to account for the window's amplitude, but we have chosen not to do this because it necessarily correlates the lowest frequencies in the data and biases the periodogram.


\section{Confusion shape variance}
\label{sec:confvartest}

\begin{figure}[h]
    \centering
    \includegraphics[width=\columnwidth]{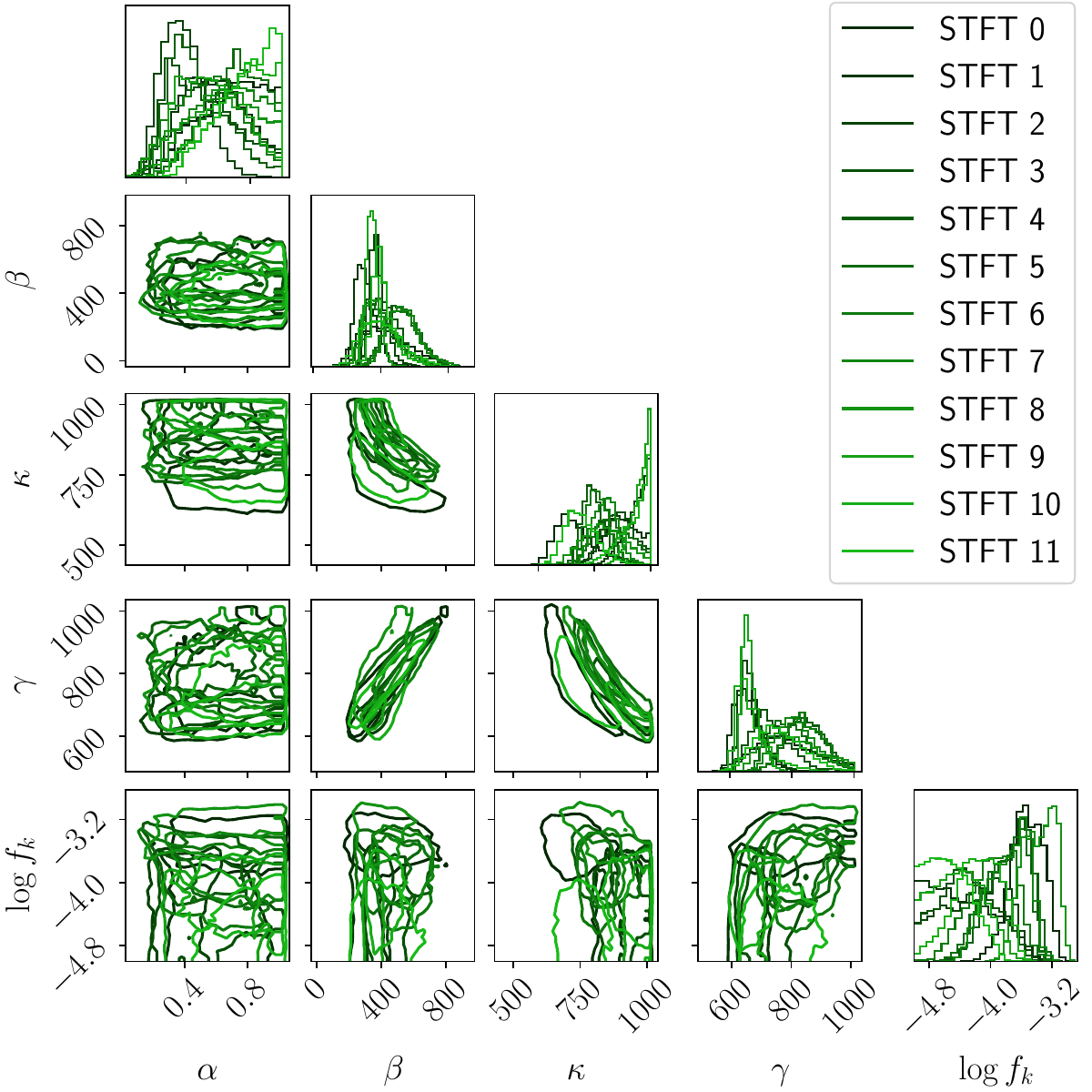}
    \caption{The confusion shape parameters when allowed to vary in each short time segment, with 12 segments total over a year of data. For clarity we plot only $2\sigma$ contours. We consider the variance of the confusion noise's shape insufficient to merit considering these parameters independent in each STFT, so we have treated them as stationary in all results quoted in this work. Also, unshown, they are not strongly correlated with any stochastic background parameters. This figure is based on the \glassviii residual with no additional signal injections, and had a total parameter space dimension of $N_{\rm s} + N_{\mathrm{STFTs}}\times N_{\rm ns} = 74$.}
    \label{fig:confvartest}
\end{figure}

\bibliography{refs}

\end{document}